\documentclass{article}
\usepackage[utf8]{inputenc}
 \usepackage[T1]{fontenc}
\usepackage{amsmath}
\usepackage{pifont}
\usepackage{bm}
 \usepackage{amssymb}
 \usepackage{amsthm}
\usepackage{parskip}
\usepackage[ruled, noend, linesnumbered]{algorithm2e}
\usepackage[colorlinks = true, pdfstartview = FitV, linkcolor = blue, citecolor = blue, urlcolor = blue, bookmarks = false]{hyperref}
\usepackage{color,xcolor}
\usepackage{setspace} 
\usepackage{graphicx}
\usepackage{subcaption}
\usepackage{setspace}
\usepackage{xcolor}
\usepackage{caption}
\usepackage{tabularray}
\UseTblrLibrary{booktabs}
\usepackage{subcaption}

\usepackage{array} 
\usepackage{amsmath} 
\usepackage{booktabs} 
\usepackage{pifont} 

\newcommand{\tick}{\ding{51}} 
\newcommand{\cross}{\ding{55}} 

\usepackage{mathtools}

\usepackage{cancel}

\usepackage{listings}
\usepackage{color} 

\usepackage[style=apa, sorting=nyt, backend=biber]{biblatex} 
\definecolor{codegreen}{rgb}{0,0.6,0}
\definecolor{codegray}{rgb}{0.5,0.5,0.5}
\definecolor{codepurple}{rgb}{0.58,0,0.82}
\definecolor{backcolour}{rgb}{0.95,0.95,0.92}

\usepackage[margin=1.25
in]{geometry}
\renewcommand{\v}[1]{\boldsymbol{#1}}

\newcommand{\ignore}[1]{}

\newsavebox\IBoxA \newsavebox\IBoxB \newlength\IHeight
\newcommand\TwoFig[6]{
  \sbox\IBoxA{\includegraphics[width=0.45\textwidth]{#1}}
  \sbox\IBoxB{\includegraphics[width=0.45\textwidth]{#4}}%
  \ifdim\ht\IBoxA>\ht\IBoxB
    \setlength\IHeight{\ht\IBoxB}%
  \else\setlength\IHeight{\ht\IBoxA}\fi
  \begin{figure}[!htb]
  \minipage[t]{0.45\textwidth}\centering
  \includegraphics[height=\IHeight]{#1}
  \caption{#2}\label{#3}
  \endminipage\hfill
  \minipage[t]{0.45\textwidth}\centering
  \includegraphics[height=\IHeight]{#4}
  \caption{#5}\label{#6}
  \endminipage 
  \end{figure}%
}

\newcommand{\ie}{\textit{i.e.}\xspace}

\makeatletter
\algocf@newcmdside@kobe{ServerComputes}{%
    \KwSty{Server} \textit{computation}%
    \ifArgumentEmpty{#1}\relax{ #1}%
    \algocf@block{#2}{end}{#3}%
    \par
}
\newcommand\Server[2]{%
    \ServerComputes{#1}%
}
\makeatother

\title{Scalable Vertical Federated Learning via Data Augmentation and Amortized Inference}
\author{Conor Hassan\footnote{\url{conorhassan.ai@gmail.com}} $\textsuperscript{ 1,2}$, Matthew Sutton$\textsuperscript{ 1,2}$, Antonietta Mira$\textsuperscript{ 3,4}$, Kerrie Mengersen$\textsuperscript{ 1,2}$ \\ 
{\small \textsuperscript{1} Centre for Data Science, Queensland University of Technology} \\ 
{\small \textsuperscript{2} School of Mathematical Sciences, Queensland University of Technology}  \\ 
{\small\textsuperscript{3} Euler Institute, Università della Svizzera italiana (USI)}\\
{\small\textsuperscript{4} Insubria University}
}

\begin{document}

\date{}
\maketitle
\begin{abstract}
Vertical federated learning (VFL) has emerged as a powerful paradigm for collaborative model estimation across multiple clients, each holding a distinct set of covariates. This paper introduces the first comprehensive framework for fitting Bayesian models in the VFL setting. We propose a novel approach that leverages data augmentation techniques to transform VFL problems into a form compatible with existing Bayesian federated learning algorithms. We present an innovative model formulation for specific VFL scenarios where the joint likelihood factorizes into a product of client-specific likelihoods. To mitigate the dimensionality challenge posed by data augmentation, which scales with the number of observations and clients, we develop a factorized amortized variational approximation that achieves scalability independent of the number of observations. We showcase the efficacy of our framework through extensive numerical experiments on logistic regression, multilevel regression, and a novel hierarchical Bayesian split NN model. Our work paves the way for privacy-preserving, decentralized Bayesian inference in vertically partitioned data scenarios, opening up new avenues for research and applications in various domains.
\end{abstract}

\section{Introduction}
\begin{refsection}
Federated learning (FL) \parencite{mcmahan2017communication, yang2019federated, kairouz2021advances} has revolutionized the field of privacy-preserving and communication-efficient learning, enabling collaborative model training across multiple \emph{clients} under the coordination of a \emph{server}, without the need for data sharing. This paradigm has unlocked new possibilities for leveraging decentralized data while maintaining strict privacy constraints, paving the way for innovative applications in various domains.

While the majority of FL algorithms focus on the \emph{horizontal} FL \parencite{yang2020horizontal} setting, where each client possesses a complete set of covariates for a unique subset of observations, the \emph{vertical} FL (VFL) \parencite{liu2019communication, wei2022vertical} setting remains relatively unexplored. In VFL, each client holds only a subset of desired covariates for all observations. This work considers two settings: (1) the response variable is shared among clients, and (2) the response variable is private to the server, and the clients send updates to the server to fit the desired statistical model.

In this paper, we aim to approximate the posterior distribution of the unknown variables given the observed data. The challenge in the VFL setting is that the likelihood function is not separable across clients, unlike in horizontal FL, where conditional independence is often assumed and exploited to create distributed or private algorithms. This non-separability arises from the fact that each client only possesses a subset of the covariates, making it impossible to compute the likelihood of any single client independently. Often, each client must contribute a value to a summation term required to evaluate the likelihood \parencite{hardy2017private, weng2020privacy, chen2021secureboost+}. Secure \emph{multi-party computation} (MPC) \parencite{cramer2015secure, li2020privacy, byrd2020differentially} is often employed to address this challenge, but it can significantly impact the scalability of the proposed methods, leaving room for further research. Moreover, to our knowledge, there is currently no literature on Bayesian modeling and inference in the VFL setting.

To overcome this challenge, we reframe the target posterior distribution via \emph{asymptotically-exact data augmentation} (AXDA) \parencite{vono2019split, vono2020asymptotically, vono2022efficient}.  This approach introduces auxiliary variables that create conditional independence between the client-specific parameters, allowing for independent updates on each client. The augmented-variable model decouples the unknown variables from the likelihood, enabling local updates on each client independently of the others. This interpretation can be seen as the probabilistic analog of an optimization problem that solves subproblems requiring information unique to each client, similar to the alternating direction method of multipliers (ADMM) \parencite{boyd2011distributed}.


Under the augmented-variable model, recently developed algorithmic schemes intended for hierarchical models in the horizontal FL setting, such as structured federated variational inference (SFVI) \parencite{hassan2023federated}, can be adapted to update the client-specific parameters. For the setting where the clients share the response variable, we develop a novel model formulation that expresses the likelihood as a product of client-specific contributions.


The auxiliary variable schemes used here require the addition of variables to the model that are equal in size to the number of observations times the number of clients. To address this, we propose a factorized amortized variational approximation, removing the increase in computational cost due to the dimensionality of the auxiliary variables per client. We provide numerical examples showcasing performance on logistic regression, multilevel regression, and a novel hierarchical Bayes split NN \parencite{poirot2019split, ceballos2020splitnn, thapa2022splitfed}.

\textbf{Contributions:} The contributions of this paper are:
\begin{enumerate}
\item We provide the first Bayesian vertical FL methods by reformulating the posterior distribution of a model in the VFL setting using data augmentation. We then use the abovementioned methods to provide the first numerical examples of fitting Bayesian regression models, including multilevel models, in the VFL setting.
\item We create a novel model formulation for the VFL setting and show its improved performance relative to existing data augmentation methods.
\item We develop a factorized amortized variational approximation that stops the inference problem scaling with the number of observations.
\item We develop a novel hierarchical Bayes split learning model.
\end{enumerate}

The paper proceeds as follows. Section \ref{section:preliminaries} provides the necessary background. Section \ref{sec::method} introduces the methods proposed. Section \ref{sec:numerics} provides numerical examples. Section \ref{sec:discussion} concludes the paper.

\section{Preliminaries}\label{section:preliminaries}

This section provides the necessary background for auxiliary variable methods and structured federated variational inference.

\subsection{Asymptotically-Exact Data Augmentation}
\label{subsec:AXDA}

In this paper, we are interested in targeting the \emph{posterior distribution} $\pi(\v\theta|\v y)\propto p(\v y|\v\theta)p(\v\theta)$ where $\v\theta\in\mathbb{R}^d$ is an unknown parameter, with prior distribution $p(\v \theta)$ and $p(\v y | \v\theta)$ is the likelihood of the observed data $\v y\in\mathbb{R}^n$. Direct inference of the posterior distribution can often be intractable or computationally expensive. \emph{Data augmentation} (DA) \parencite{van2001art} is a common approach to address this issue by introducing auxiliary variables $\v z \in \mathbb{R}^m$ into the statistical model through a model-specific joint density $\pi(\v \theta, \v z | \v y)$. This joint density is chosen to be simpler to compute and satisfies the marginalization property, 
\begin{align}
\pi(\v\theta|\v y)=\int \pi(\boldsymbol{\theta},\v{z}|\v y)\mathrm{d}\v{z}.  \label{eq:da_target_density}
\end{align}
\emph{Asymptotically-exact data augmentation} (AXDA) \parencite{vono2020asymptotically} is a flexible form of DA that relaxes the marginalization property in \eqref{eq:da_target_density} by introducing an augmented distribution $p(\v z | \v \theta; \rho)$ that depends on a positive scalar hyperparameter $\rho > 0$. The resulting augmented posterior distribution is given by
\begin{align}
\pi(\v\theta|\v y; \rho) = \int \pi(\v \theta,\v{z}| \v y; \rho)\mathrm{d}\v{z},
\label{eq:adxa_target_density}
\end{align}
where $\pi(\v\theta, \v z |\v y; \rho)\propto p(\v y| \v z, \v\theta)p(\v z,\v \theta; \rho)$, with augmented likelihood $p(\v y | \v z, \v\theta)$ and prior $p(\v z,\v \theta; \rho)=p(\v z|\v \theta; \rho)p(\v \theta)$. The conditional distribution $p(\v z|\v \theta; \rho)$ is chosen to be a Gaussian with mean equal to a function of $\v \theta$ and variance $\rho^2$. The key property of AXDA is that, under the assumption that the mean function of $\v z|\v \theta; \rho$ is chosen correctly, the augmented posterior distribution $\pi(\v \theta | \v y; \rho)$ converges to the original posterior distribution $\pi(\v \theta| \v y)$ as $\rho \rightarrow 0$, \ie $\lim_{\rho \rightarrow 0} \pi(\v\theta | \v y; \rho) = \pi(\v\theta| \v y)$. This property justifies the name \emph{``asymptotically-exact''} and ensures that the AXDA framework can provide a valid approximation to the original posterior distribution. AXDA offers several advantages over traditional data augmentation methods. It provides a flexible framework that can be applied to a wide range of models without requiring model-specific choices. Additionally, the introduction of the hyperparameter $\rho$ allows for a trade-off between computational tractability and the accuracy of the approximation.

\subsection{Structured Federated Variational Inference}\label{sec:SFVI_prelim}
Variational inference (VI) \parencite{blei2017variational, zhang2018advances} is a powerful algorithmic framework for approximating intractable posterior distributions $\pi(\v\theta|\v y)$. The core idea behind VI is to introduce a family of tractable distributions $q_{\v\phi}(\v\theta)$, parameterized by variational parameters $\v\phi$, and find the member of this family that best approximates the true posterior.

The optimal variational approximation is found by minimizing the Kullback-Leibler (KL) divergence between $q_{\v\phi}(\v\theta)$ and $\pi(\v\theta|\v y)$, which is equivalent to maximizing the evidence lower bound (ELBO), 
\begin{align}
\mathcal{L}(\v\phi) = \mathbb{E}_{\v\theta \sim q_{\v\phi}}[\log p(\v y, \v\theta) - \log q_{\v\phi}(\v\theta)].
\end{align}
The ELBO is a lower bound on the log marginal likelihood $\log p(\v y)$ and can be used as a surrogate objective function for optimization. If the variational approximation $q_{\v\phi}(\v\theta)$ exactly recovers the true posterior $\pi(\v\theta|\v y)$, the ELBO is equal to the log marginal likelihood.

To optimize the ELBO, VI methods typically employ gradient-based optimization algorithms \parencite{ruder2017overview}. A key challenge in VI is to obtain low-variance, unbiased gradient estimates of the ELBO with respect to the variational parameters $\v\phi$. The reparameterization trick \parencite{kingma2013auto, rezende2014stochastic} has emerged as a powerful technique for addressing this challenge, which allows for an efficient and potentially low-variance estimate of the ELBO. The reparameterization involves expressing a sample of $\v\theta$ drawn from $q_{\v\phi}(\v\theta)$ as a deterministic function of $\v\phi$ by transforming random noise $\v\epsilon$: $\v\theta = f_{\v\phi}(\v \epsilon)$. The noise variable $\v\epsilon$ is drawn from a distribution $q_{\v\epsilon}$, which is independent of $\v\phi$. In this context, $f_{\v\phi}$ might be, for example, a location-scale transform. Given this formulation, a resulting single-sample unbiased Monte Carlo estimator of the desired gradient vector is the \emph{sticking-the-landing} (STL) estimator \parencite{roeder2017sticking},
\begin{align}
\hat{\nabla}_{\v\phi}\mathcal{L} := {\frac{\partial f_{\v\phi}(\v\epsilon)}{\partial\v\phi}}^\top \nabla_{\v \theta}[\log p(\v \theta, \v y) - \log q_{\v\phi}(\v \theta)].
\end{align}

Structured Federated Variational Inference (SFVI) \parencite{hassan2023federated} is a framework that uses structured approximations \parencite{hoffman2015stochastic, ranganath2016hierarchical, tan2020conditionally} to fit hierarchical Bayesian models in the horizontal FL setting. The models considered have a joint probability distribution of the form, 
\begin{align*}
p(\v y, \v z, \v\theta) = p(\v z)\prod_{j=1}^J p(\v\theta_j|\v z)p(\v y | \v\theta_j, \v z),
\end{align*}
where $\v\theta = (\v\theta_1^\top, \ldots, \v\theta_J^\top)^\top$ such that the parameters $\v\theta_{j}$ only appear in the likelihood contribution from the $j$-th client and are conditionally independent across clients, while $\v z$ are shared global parameters that appear in all likelihood contributions across clients. SFVI proposes to use a factorized variational approximation, 
\begin{align*}
q_{\v\phi}(\v z, \v\theta) = q_{\v\phi_{\v z}}(\v z)\prod_{j=1}
^J q_{\v\phi_j}(\v \theta_{j}|\v z),\end{align*}
where $q_{\v\phi_z}(\v z)$ is Gaussian distributed, parameterized by $\v \phi_z$, 
and each $q_{\v\phi_j}(\v \theta_{j}|\v z)$ is Gaussian distributed, parameterized by $\v \phi_j$ respectively, and potentially include dependencies on $\v z$. 

\cite{hassan2023federated} derive two algorithms, SFVI and SFVI-Avg, for combinations of models and algorithms that share this form. Both of these algorithms enable the updating of the local variational parameters $\v\phi_j$ for each client in parallel and \emph{privately} (\ie they are not shared with the server), only requiring communicating gradient information regarding $\v\phi_z$ to the server at each iteration.

The SFVI framework provides an effective way to apply VI in the horizontal FL setting, allowing for efficient and privacy-preserving model fitting. In the following sections, we will discuss how the concepts of AXDA and SFVI can be leveraged to develop methods for the vertical FL setting.


\section{Auxiliary Variable Methods for Vertical Federated Learning}
\label{sec::method}
This section introduces \emph{asymptotically-exact data augmentation} (AXDA), similar in flavor to that of \cite{vono2019split, vono2020asymptotically, vono2022efficient, rendell2020global}, for vertical FL. This scheme incorporates \emph{auxiliary variables} based on the \emph{features} (e.g., covariates), facilitating the construction of a VFL algorithm for specific classes of statistical models. We then present an alternative modeling approach for specific vertical FL settings. As both the augmented-variable model (presented in Section \ref{subsec:augmented_model}) and the power-likelihood model (presented in Section \ref{subsec:power_likelihood}) satisfy the same asymptotically-exact property as AXDA, the theory presented for AXDA is equally applicable for the two following models presented.

\subsection{Vertical Federated Learning Motivation and Setting}
\label{subsec:motivation}

We aim to perform inference for the unknown variable of interest $\v \theta \in \mathbb{R}^p$. In the vertical federated learning (VFL) setting, the joint data matrix $\v x \in \mathbb{R}^{n \times p}$ is distributed across $J$ clients, such that it can be written as $\v x = (\v x_1, \ldots, \v x_J)$, where each $\v x_j \in \mathbb{R}^{n \times p_j}$ and $\sum_{j=1}^J p_j = p$. This distribution of the data matrix occurs because each client holds a subset of the features for all observations, necessitating collaboration among clients to perform inference on the complete set of features. We also assume that the parameter vector $\v\theta$ is distributed across clients as $\v\theta=(\v\theta_1^\top, \ldots, \v\theta_J^\top)^\top$. In this paper, we aim to approximate models of the form
\begin{align}
\pi(\v\theta|\v y)\propto \exp\left(\log p\bigg (\v y \bigg | \sum_{j=1}^J g_j(\v x_j, \v \theta), \v\gamma
\bigg )+\sum_{j=1}^J \log p_j(\v \theta_j)\right),
\end{align}
in the VFL setting, where each client $j$ has the choice to set their own prior distribution $p(\v\theta_j):\mathbb{R}^{p_j}\rightarrow \mathbb{R}$, for the parameters $\v\theta_j$ associated with their subset of the data matrix $\v x_j$, and $\v\gamma$ is an unknown parameter shared across the clients. The challenge arises in evaluating the likelihood $p(\v y | \sum_{j=1}^J g_j(\v x_j, \v \theta), \v\gamma)$ in the VFL setting because each client only has access to a subset of the data and parameters, making it impossible to compute the likelihood independently.\\
\emph{Running example:} Consider linear regression with unknown mean and variance, where $J$ clients hold the desired covariates. By setting $\v \theta_j = \v \beta_j$ for all $J$ clients, $\v \gamma = \sigma^2$, and $g_j(\v x, \v \theta) = \v x_j^\top \v \theta_j$, the likelihood $f$ can be can be expressed as $\v y|\v \beta, \sigma^2 \sim \mathcal{N}(\sum_{j=1}^J \v x_j^\top\v \beta_j, \sigma^2)$.

Inference of such a target density is difficult in the VFL setting because each client $j$ would be required to share the value of their likelihood contribution $g_j(\v x, \v\theta)$. Thus, it would not be possible to perform multiple computation steps updating $\v\theta_j$ on client $j$ without communicating with the server. This issue motivates us to formulate the target density in an alternative manner to enable the use of an existing FL algorithm that can handle the unique challenges posed by the VFL setting while maintaining privacy.

\subsection{Augmented-Variable Model}\label{subsec:augmented_model}
Here, we present an AXDA setting suitable for VFL. Let $\v z =(\v z_1^\top, \ldots, \v z_J^\top)^\top$, where $\v z_j\in\mathbb{R}^n$ denotes the \emph{auxiliary variables} introduced with respect to client $j$. The joint posterior of $\v\theta$ and $\v z$ in the \emph{augmented-variable model} is proportional to
\begin{align}
\pi(\v\theta, \v z | \v y; \rho)\propto \exp\left(\log p(\v y| \v z, \v \gamma) + \sum_{j = 1}^J \log p(\v z_j | \v\theta_j; \rho) + \sum_{j=1}^J \log p(\v\theta_j)\right),
\end{align}
where $p(\v y | \v z, \v \gamma) := p(\v y|\sum_{j=1}^J g_j(\v x_j, \v \theta_j), \v \gamma)$ is the likelihood, with each $g_j(\v x_j, \v\theta_j) = \v z_j$. This formulation effectively decouples the likelihood from the original parameters $\v \theta$, making it a function of the auxiliary variables $\v z$.

The conditional distribution $\v z_j | \v\theta_j; \rho$ is chosen to be a Gaussian distribution with mean equal to $g_j(\v x_j, \v \theta_j)$, and variance $\rho^2$. This choice of conditional distribution is motivated by the AXDA framework, which allows for a flexible and computationally tractable augmented posterior distribution. The conditional distribution of $\v\theta_j$ is proportional to $p(\v\theta_j | \v\theta_{-j}, \v z; \rho) \propto p(\v z_j | \v \theta_j; \rho)p(\v\theta_j)$, depending only on the auxiliary variable $\v z_j$ and the prior $p_j(\v\theta_j)$.\\
\emph{Running example:} Consider a linear regression with unknown mean and variance, where $J$ clients hold the desired covariates. Setting $\v\gamma=\sigma^2$ and $\v\theta_j=\v\beta_j$, $g_j(\v x_j, \v\theta_j)=\v x_j^\top \theta_j$, 
for all $J$ clients, 
the augmented-variable model can be written as
\begin{align*}
\v y | \v z &\sim \mathcal{N}\big (\sum_{j=1}^J \v z_j, \sigma \big ), \\
\v z_j | \v\theta_j &\sim \mathcal{N}(\v x_j^\top\v \beta_j, \rho), \quad j = 1, \ldots, J, \\
\v\theta_j &\sim p(\v\theta_j), \quad j = 1, \ldots, J.
\end{align*}

As in AXDA, the augmented posterior distribution $\pi(\v \theta| \v y; \rho)$ of this model converges to the original posterior distribution $\pi(\v \theta| \v y)$ as $\rho \rightarrow 0$, \ie $\lim_{\rho \rightarrow 0} \pi(\v\theta| \v y; \rho) = \pi(\v\theta| \v y)$. Given the formulation of the augmented-variable model, it follows that, for all pairs of clients $i, k$, we have the conditional independence property $\v \theta_i \perp\!\!\!\perp  \v\theta_k | \v z_i, \v z_k$, meaning that the elements of $\v\theta$ can be updated in parallel.

In addition to satisfying the above conditional independence property, the augmented-variable model can be fit in two FL scenarios:

\begin{enumerate}
\item The response variable $\v y$ is shared across clients; clients can either update their respective $\v z_j$ and send it to the server or send the necessary gradient to the server, where the server carries out the update; the server is responsible for redistributing updated values of $\v z$ among clients.
\item The response variable $\v y$ is not shared and is only available to the server; the clients send the server the necessary gradients to update their $\v z_j$; the server updates all $\v z$ and redistributes to the clients.
\end{enumerate}

\subsection{Power-Likelihood Model}\label{subsec:power_likelihood}

The likelihood contribution $g_j(\v \theta_j, \v x_j)$ is known locally to client $j$ in the VFL setting. This allows for an alternative formulation of the target density, where the likelihood evaluated at client $j$ is a function of $g_j(\v x_j, \v \theta_j)$ rather than the auxiliary variable $\v z_j$, as in the augmented-variable model. This leads to a log-likelihood that is a weighted summation of $J$ client-specific log-likelihood contributions. We write the target density as

\begin{align}
\pi(\v\theta, \v z | \v y; \rho)\propto \exp\left(\frac{1}{J}\sum_{j=1}^J \log p(\v y| \v g_j(\v \theta_j, \v x_j), \v z_{-j}) + \sum_{j=1}^J \log p(\v\theta_j)+ \sum_{j=1}^J \log p(\v z_j|\v\theta;\rho)\right).\label{eqn:power_joint_target}
\end{align}

It is important to note that the likelihood in the power-likelihood model has a different form compared to the augmented-variable model. In the power-likelihood model, the likelihood is a product of $J$ client-specific likelihood contributions, each raised to the power of $1/J$. This formulation differs from the general framework presented in Section \ref{subsec:augmented_model}, where the likelihood is a function of the sum of the $g_j$ functions.\\
\emph{Running example:} Consider the linear regression with unknown mean and variance. The \emph{power-likelihood model} can be written as

\begin{align*}
\v y | \v \theta, \v z &\sim \prod_{j=1}^J \mathcal{N}\bigg (\v x_j^\top\v\beta_j + \sum_{k\neq j}^J \v z_k, \sigma \bigg )^{1/J}, \\
\v z_j | \v\theta_j &\sim \mathcal{N}(\v x_j^\top\v \beta_j, \rho), \quad j = 1, \ldots, J, \\
\v\theta_j &\sim p(\v\theta_j), \quad j = 1, \ldots, J.
\end{align*}

Despite the difference in the likelihood formulation, the power-likelihood model still maintains the AXDA property, where the augmented posterior distribution converges to the original posterior distribution as $\rho \rightarrow 0$, \ie $\lim_{\rho \rightarrow 0} \pi(\v\theta | \v y; \rho) = \pi(\v\theta| \v y)$.

Unlike the augmented-variable model, the power-likelihood model can only be fit in the setting where the response variable $\v y$ is known to the clients, and the clients share the necessary gradients with the server to update their respective variables $\v z_j$. This limitation arises because the likelihood of the power-likelihood model has an \emph{explicit} dependence on the data $\v x$ and the parameters $\v\theta$.

\begin{figure}[!ht]
\centering
\includegraphics[width=0.96\textwidth]{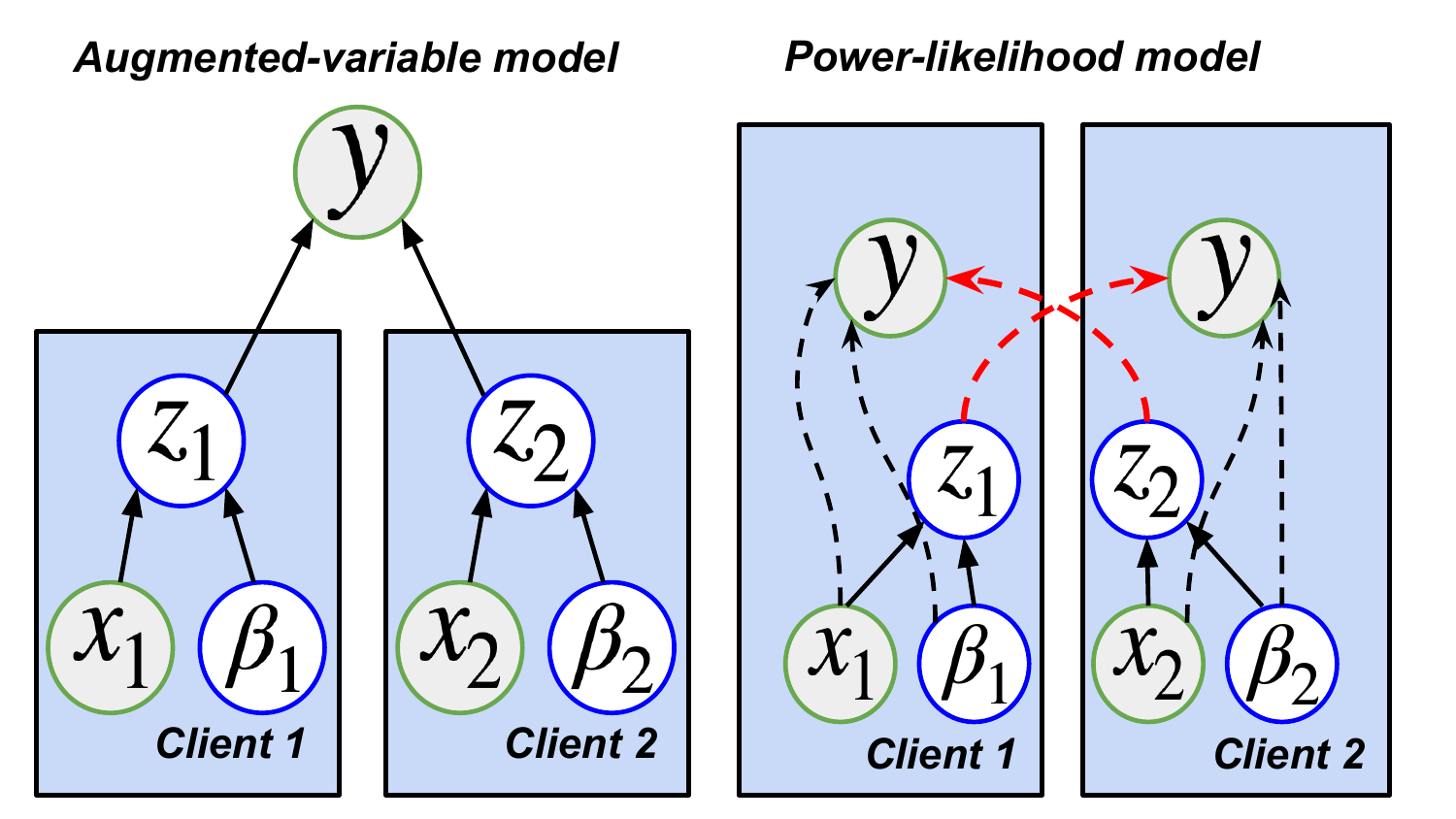}
\caption{Graphical illustration of the dependency structure of the proposed models: \emph{augmented-variable} (left), \emph{power-likelihood} (right)}
\label{fig:model-structure}
\end{figure}

Figure \ref{fig:model-structure} graphically illustrates the flow of information and the dependency structure required for the \emph{augmented-variable} and the \emph{power-likelihood} models, respectively.

The power-likelihood model provides an alternative formulation for the VFL setting that prioritizes improved modeling by allowing the likelihood evaluated at each client to be a function of their local likelihood contribution. However, this comes at the cost of reduced flexibility in the settings where the model can be fit, as it requires the clients to know the response variable. In contrast, the augmented-variable model offers more flexibility regarding the settings where it can be applied, albeit with a different likelihood formulation.

\section{Algorithm details}

This section describes how the augmented-variable model in Section \ref{subsec:augmented_model} and the power-likelihood model in Section \ref{subsec:power_likelihood} can be inferred in the VFL setting. Typically, one would estimate the posterior distribution $\pi(\v \theta, \v z | \v y; \rho)$ using \emph{Markov chain Monte Carlo} (MCMC) methods \parencite{brooks2011handbook}. However, updating the auxiliary variables specific to each client in the VFL setting requires the conditional distribution $\v z_j | \v{z}_{-j}, \v\theta_j$. Assuming that we want to reduce communication of data-related quantities, such as gradients of log densities, from clients to the server, we use alternative algorithms that can exploit the conditional independence structure among the clients given $\v z$.

Two such algorithms are \emph{structured federated variational inference} (SFVI) \parencite{hassan2023federated} and the \emph{stochastic optimization via unadjusted Langevin} (SOUL) \parencite{de2021efficient, kotelevskii2022fedpop} algorithm. These algorithms leverage the conditional independence structure by allowing for local updates on each client and reducing the need for frequent communication between the clients and the server. In this section, we will focus on describing SFVI in detail, while the discussion of SOUL is provided in Appendix \ref{appendix:SOUL}. 

We will explore two variational approximations suitable for the FL setting:
\begin{enumerate}
    \item A \emph{structured variational approximation} that employs ideas from amortized inference. This approximation leverages the conditional independence structure of the model and enables efficient local updates on each client.
    \item A \emph{mean-field variational approximation}, which assumes independence among the variables in the approximation.
\end{enumerate}

\subsection{Amortized Structured Variational Approximation}\label{sec:amortized_variational_family}
For the structured variational approximation, we choose a similar form to \cite{hassan2023federated}, such that
\begin{align*}
q_{\v\phi, \v\psi}(\v \theta, \v z) = \prod_{j=1}^J q_{\v\phi_j}(\v\theta_j)q_{\v\psi_j}(\v z_j | \v\theta_j),
\end{align*}
where $q_{\v\phi_j}(\v\theta_j)$ and $q_{\v\psi_j}(\v z_j | \v\theta_j)$ are the variational distributions for the parameters $\v \theta_j$ and auxiliary variables $\v z_j$ of client $j$, parameterized by variational parameters $\v \phi_j$ and $\v \psi_j$ respectively. 

For the parameterization of the variational approximation of $\v\theta_j$, we choose that $
q_{\v\phi_j}(\v\theta_j)=\mathcal{N}(\v\mu_{\v\theta_j}, \v\Sigma_{\v\theta_j})$, where $\v\phi_j=(\v\mu_{\v\theta_j}^\top,\text{vec}(\v\Sigma_{\v\theta_j})^\top)^\top$ with the reparameterization applied such that $\v\theta_j=\v\mu_{\v\theta_j} +\v\Sigma_{\v\theta_j} \v\epsilon_j$. For the parameterization of the variational approximation of $\v z_j$, we propose different parameterizations for the augmented-variable and power-likelihood models. In the augmented-variable model, we choose $q_{\v\psi_j}(\v z_j | \v\theta_j)=\mathcal{N}\big (\v\mu_j( \v x_j^\top\v\theta_j), \v\sigma_j(\v x_j^\top \v\theta_j)\big )$. For the power-likelihood model, we choose $q_{\v\psi_j}(\v z_j | \v\theta_j)=\mathcal{N}\big (\v\mu_j(\v y, \v x_j^\top\v\theta_j), \v\sigma_j(\v y, \v x_j^\top \v\theta_j)\big )$. In both cases, $\v \mu_j(\cdot)$ and $\v \sigma_j(\cdot)$ are output by a NN parameterized by $\v\psi_j$ for $j=1, \ldots, J$. Using a neural network allows for a flexible and expressive parameterization of the variational approximation for $\v z_j$, enabling it to capture complex dependencies between $\v z_j$ and $\v \theta_j$.

As a consequence of the choice of factorization, 
we apply the appropriate joint reparameterization, 
\begin{align}
\v\theta_j &= \v \mu_{\v\theta_j} + \v\Sigma_{\v\theta_j}\v\epsilon_j, \quad j=1, \ldots, J,\label{eqn:f_reparam}\\
\v z_j &= \v\mu_j(\v x_j^\top\v\theta_j) + \v\sigma_j(\v x_j^\top\v\theta_j) \cdot \v\tau_j, \quad j=1, \ldots, J\label{eqn:g_reparam},
\end{align}
where the vectors $\big \{(\v\tau_j, \v\epsilon_j)\big \}_{j=1}^J$ are drawn from a standard Gaussian distribution. 

The amortized structured variational approximation is efficient because the cost of representing $q_{\v\psi_j}(\v z_j | \v\theta_j)$ scales with the size of the neural network parameterization $\v\psi_j$, rather than the dimension of $\v z_j$ or the number of observations. The variational approximation's computational cost and memory footprint remain manageable even when dealing with high-dimensional auxiliary variables or large datasets, making it suitable for a wide range of practical applications.

\subsection{Mean-field Variational Approximation}\label{sec:MFVI}
The mean-field approximation that we parameterize takes the form 
\begin{align*}
q_{\v\phi, \v\psi}(\v \theta, \v z) = \prod_{j=1}^J q_{\v\phi_j}(\v\theta_j)q_{\v\psi_j}(\v z_j),
\end{align*}
where the only difference with Section \ref{sec:amortized_variational_family} is that we choose $q_{\v\psi_j} (\v z_j)$ such that $q_{\v\psi_j} (\v z_j)=\mathcal{N}(\v \mu_{\v z_j}, \v \sigma_{\v z_j})$ where $\v\psi_j = (\v\mu_{\v z_j}^\top, \v\sigma_{\v z_j}^\top)^\top$ such that the variational approximation for $\v z_j$ no longer has dependence on $\v\theta_j$. This form of variational approximation simplifies the reparameterization, 
\begin{align}\label{eqn:mfvi_reparam}
\v z_j = \v\mu_{\v z_j} + \v \sigma_{\v z_j} \cdot \v\tau_j, \quad j=1, \ldots, J,
\end{align}
where $g$ is now only parameterized by $\v\psi_j$ rather than both $\v\psi_j$ and $\v\phi_j$.

\subsection{Algorithm details for Augmented-variable model}\label{sec:SFVI_augmented}
This section describes the algorithm steps required to fit the augmented-variable model using the structured or mean-field variational approximation. 
We begin by considering the evidence lower bound (ELBO) of the augmented-variable model using the structured variational approximation, 
\begin{align*}
\mathcal{L}&=\mathbb{E}_{\prod_{j=1}^J q(\v z_j | \v\theta_j)q(\v\theta_j)}\bigg [\underbrace{\log p(\v y|\v z )}_{\textcolor{red}{\text{dependent}}} + \sum_{j=1}^J \bigg [\underbrace{\log \frac{p(\v z_j|\v\theta_j)}{ q_{\v\psi_j}(\v z_j|\v \theta_j)}+\log \frac{p(\v\theta_j)}{q_{\v\phi_j}(\v\theta_j)}}_{\textcolor{blue}{\text{independent}}}\bigg ] \bigg ]\\ 
&= \mathbb{E}_{\prod_{j=1}^J q(\v z_j | \v\theta_j)}\bigg [\log p(\v y|\v z )\bigg ] + \sum_{j=1}^J \mathbb{E}_{q(\v z_j | \v\theta_j)q(\v\theta_j)}\bigg [ \log \frac{ p(\v z_j|\v\theta_j)}{ q_{\v\psi_j}(\v z_j|\v \theta_j)}+\log \frac{ p(\v\theta_j)}{ q_{\v\phi_j}(\v\theta_j)}\bigg ]\\
&= \textcolor{red}{\mathcal{L}_0} + \sum_{j=1}^J  \textcolor{blue}{\mathcal{L}_j}.
\end{align*}
The ELBO decomposes into a  term, $\mathcal{L}_0$ (in \textcolor{red}{red}), which involves the likelihood $p(\v y|\v z)$ and requires global information depending on all the parameters and auxiliary variables, and an independent terms, $\mathcal{L}_j$ (in \textcolor{blue}{blue}), which can be computed locally by each client. Due to the structure in both the model and the variational approximation, the following gradients are equal to 
\[\frac{\partial \mathcal{L}}{\partial\v\theta}_j= \frac{\partial \mathcal{L}_j}{\partial\v\theta_j} \quad\quad \text{and} \quad\quad \frac{\partial \mathcal{L}}{\partial \v z_j} = \frac{\partial\mathcal{L}_0}{\partial\v z_j} + \frac{\partial\mathcal{L}_j}{\partial\v z_j}.\]

Exploiting the sparsity in the Jacobian matrix of the reparameterization, the gradient estimate of $\widehat{\nabla}_{\v\phi_j}\mathcal{L}$ can be expressed as
\begin{align}
\widehat{\nabla}_{\phi_j}\mathcal{L} &= \frac{\partial f(\v\epsilon_j)^\top}{\partial \v\phi_j}\frac{\partial\mathcal{L}}{\partial\v\theta_j} + \frac{\partial g(\v\tau_j, \v\epsilon_j)^\top}{\partial\v\phi_j} \frac{\partial\mathcal{L}}{\partial\v z_j}, \nonumber\\ 
&= \textcolor{blue}{\frac{\partial f(\v\epsilon_j)^\top}{\partial \v\phi_j}\frac{\partial\mathcal{L}_j}{\partial\v\theta_j}} + \textcolor{blue}{\frac{\partial g(\v\tau_j, \v\epsilon_j)^\top}{\partial\v\psi_j}} \bigg [\textcolor{red}{\frac{\partial\mathcal{L}_0}{\partial\v z_j}} + \textcolor{blue}{\frac{\partial\mathcal{L}_j}{\partial\v z_j}} \bigg ]\label{eqn:augmented_phi_gradient}, 
\end{align}
and the gradient $\widehat{\nabla}_{\v\psi_j} \mathcal{L}$ is given by
\begin{align}
\widehat{\nabla}_{\v\psi_j} \mathcal{L}&=  \frac{\partial g(\v\tau_j, \v\epsilon_j)^\top}{\partial\v\psi_j} \frac{\partial\mathcal{L}}{\partial\v z_j}, \nonumber\\
&= \textcolor{blue}{\frac{\partial g(\v\tau_j, \v\epsilon_j)^\top}{\partial\v\phi_j}} \bigg [\textcolor{red}{\frac{\partial\mathcal{L}_0}{\partial\v z_j}} + \textcolor{blue}{\frac{\partial\mathcal{L}_j}{\partial\v z_j}} \bigg ]\label{eqn:augmented_psi_gradient}, 
\end{align}
where, following the same color coding introduced above, terms in \textcolor{blue}{blue} can be computed independently by each client, and terms in \textcolor{red}{red} must be computed on the server. 

When using the mean-field variational approximation, the dependency of the auxiliary variables $\v z_j$ on the variational parameters $\v \phi_j$ vanishes, resulting in a simplified gradient for $\widehat{\nabla}_{\v\phi_j}\mathcal{L}$, 
\begin{align}
\widehat{\nabla}_{\phi_j}\mathcal{L} &= \frac{\partial f(\v\epsilon_j)^\top}{\partial \v\phi_j}\frac{\partial\mathcal{L}}{\partial\v\theta_j} + \underbrace{\cancel{\frac{\partial g(\v\tau_j, \v\epsilon_j)^\top}{\partial\v\phi_j} \frac{\partial\mathcal{L}}{\partial\v z_j}}}_{\text{dependency removed}}, \nonumber\\ 
&= \textcolor{blue}{\frac{\partial f(\v\epsilon_j)^\top}{\partial \v\phi_j}\frac{\partial\mathcal{L}_j}{\partial\v\theta_j}}\label{eqn:augmented_phi_mfvi_gradient}. 
\end{align}
Having derived the necessary gradients for updating the variational parameters $\v\phi_j$ and $\v\psi_j$, we now present Algorithm \ref{alg:augmented_variable} for fitting the augmented-variable model using either the amortized structured or the mean-field variational approximation. The algorithm follows an iterative scheme where, at each iteration, the server calculates the gradient of the likelihood with respect to the auxiliary variables and sends the respective gradients to the clients. The clients then compute the required terms to update their parameters $\v\theta_j$ and auxiliary variables $\v z_j$ locally and return the updated auxiliary variables to the server. This process is repeated for a specified number of iterations or until convergence.

\begin{algorithm}[ht]
\SetAlgoLined
\KwIn{\textit{Server}: number of iterations $N$. \textit{Clients}: initial local variational parameters $\v\phi_j, \v\psi_j$.}
\KwOut{\textit{Clients}: variational parameters $\v\phi_j, \v\psi_j$.}
\textit{Initialization step:} Each client generates their value for $\v z_j$ and sends it to the server\\
\For{$i=1, \ldots, N$}{
\Server{
Receive $\v z_j$ from client $j=1, \ldots, J$\\
Calculate ${\frac{\partial\mathcal{L}_0}{\partial \v z_j}}$ for $j=1, \ldots, J$\\
Send ${\frac{\partial\mathcal{L}_0}{\partial \v z_j}}$ to client $j$ for $j = 1, \ldots, J$

}

\For{each client $j$ in parallel}{
Receive ${\frac{\partial\mathcal{L}_0}{\partial \v z_j}}$ from the server\\
Generate $\v\theta_j$ via \eqref{eqn:f_reparam}\\
Calculate $\widehat{\nabla}_{\v\phi_j}\mathcal{L}$ via \eqref{eqn:augmented_phi_gradient} for structured or \eqref{eqn:augmented_phi_mfvi_gradient} for mean-field\\
Calculate $\widehat{\nabla}_{\v\psi_j}\mathcal{L}$ via \eqref{eqn:augmented_psi_gradient}\\
$\v\phi_j \gets {\sf optimizer.step}\big (\widehat{\nabla}_{\v\phi_j}\mathcal{L}\big )$\\
$\v\psi_j \gets {\sf optimizer.step}\big (\widehat{\nabla}_{\v\psi_j}\mathcal{L}\big )$\\
Generate $\v z_j$ via \eqref{eqn:g_reparam} for structured or \eqref{eqn:mfvi_reparam} for mean-field\\
Send $\v z_j$ to server}

}
\caption{SFVI for auxiliary-variable model}\label{alg:augmented_variable}
\end{algorithm}

\subsection{Algorithm details for Power-likelihood Model}\label{sec:SFVI_power}
The ELBO of the power-likelihood model with the amortized structured variational approximation is
\begin{align}
\mathcal{L}&=\mathbb{E}_{\prod_{j=1}^J q_{\v\psi_j}(\v z_j | \v\theta_j)q_{\v\phi_j}(\v\theta_j)}\bigg [\frac{1}{J}\sum_{j=1}^J \underbrace{\log p(\v y|\v\theta_j, \v z_{-j} )}_{\textcolor{red}{\text{dependent}}} + \sum_{j=1}^J \bigg [\underbrace{\log \frac{p(\v z_j|\v\theta_j)}{q_{\v\psi_j}(\v z_j|\v \theta_j)}+\log \frac{ p(\v\theta_j)}{q_{\v\phi_j}(\v\theta_j)}}_{\textcolor{blue}{\text{independent}}}\bigg ] \bigg ]\nonumber\\ 
&= \sum_{j=1}^J \frac{1}{J}\mathbb{E}_{q_{\v\phi_j}(\v\theta_j)\prod_{k\neq j}^J q_{\v\psi_k}(\v z_j | \v\theta_j)}\bigg [\log p(\v y|\v\theta_j, \v z_{-j} )\bigg ] \\
&\qquad\qquad + \sum_{j=1}^J \mathbb{E}_{q_{\v\psi_j}(\v z_j | \v\theta_j)q_{\v\phi_j}(\v\theta_j)}\bigg [ \log \frac{p(\v z_j|\v\theta_j)}{q_{\v\psi_j}(\v z_j|\v \theta_j)}+\log \frac{ p(\v\theta_j)}{ q_{\v\phi_j}(\v\theta_j)}\bigg ]\nonumber\\
&= \sum_{j=1}^J \textcolor{red}{\mathcal{L}_{0, j}} + \sum_{j=1}^J  \textcolor{blue}{\mathcal{L}_{1, j}}\nonumber.
\end{align}
Due to the dependence structure in the ELBO, the following gradients are equal to
\[\frac{\partial \mathcal{L}}{\partial\v\theta}_j= \frac{\partial \mathcal{L}_{0, j}}{\partial\v\theta_j}\ + \frac{\partial\mathcal{L}_{1, j}}{\partial \v\theta_j}\] and \[\frac{\partial \mathcal{L}}{\partial \v z_j} = \frac{\partial\mathcal{L}_{1, j}}{\partial\v z_j} + \sum_{k\neq j}^J \bigg [\frac{\partial\mathcal{L}_{0,k}}{\partial\v z_j} \bigg ].\]
Due to the sparsity in the Jacobian matrix  of the reparameterization used by the amortized structured variational approximation, the gradient of the ELBO with respect to the variational parameters $\v\phi_j$ can be written as 
\begin{align}
\widehat{\nabla}_{\v\phi_j}\mathcal{L} &= \frac{\partial f(\v\epsilon_j)^\top}{\partial \v\phi_j}\frac{\partial\mathcal{L}}{\partial\v\theta_j} + \frac{\partial g(\v\tau_j, \v\epsilon_j)^\top}{\partial\v\phi_j} \frac{\partial\mathcal{L}}{\partial\v z_j}, \nonumber\\ 
&= \textcolor{blue}{\frac{\partial f(\v\epsilon_j)^\top}{\partial \v\phi_j}}\bigg [\textcolor{red}{\frac{\partial\mathcal{L}_{0,j}}{\partial\v\theta_j}} + \textcolor{blue}{\frac{\partial\mathcal{L}_{1,j}}{\partial\v\theta_j}} \bigg ]+ \textcolor{blue}{\frac{\partial g(\v\tau_j, \v\epsilon_j)^\top}{\partial\v\phi_j}} \bigg [\textcolor{red}{\sum_{k\neq j}^J\bigg [\frac{\partial\mathcal{L}_{0, k}}{\partial\v z_j}\bigg ]} + \textcolor{blue}{\frac{\partial\mathcal{L}_{1, j}}{\partial\v z_j}} \bigg], \label{eqn:power_phi_gradient}
\end{align} 
where terms in blue can be computed on a particular client with no added information from other clients required, and terms in red mean that information from other clients is required to calculate this term. The gradient of the ELBO with respect to the variational parameters $\v\psi_j$ are
\begin{align}
\widehat{\nabla}_{\v\psi_j}\mathcal{L} &=
\frac{\partial g(\v\tau_j, \v\epsilon_j)^\top}{\partial\v\psi_j} \frac{\partial\mathcal{L}}{\partial\v z_j}, \nonumber\\ 
&=\textcolor{blue}{\frac{\partial g(\v\tau_j, \v\epsilon_j)^\top}{\partial\v\psi_j}} \bigg [\textcolor{red}{\sum_{k\neq j}^J\bigg [\frac{\partial\mathcal{L}_{0, k}}{\partial\v z_j}\bigg ]} + \textcolor{blue}{\frac{\partial\mathcal{L}_{1, j}}{\partial\v z_j}} \bigg ]\label{eqn:power_psi_gradient}.
\end{align}
Similar to the mean-field approximation with the auxiliary-variable model, when using the mean-field approximation with the power-likelihood model, the gradient of the ELBO with respect to $\v\phi_j$ simplifies to 
\begin{align}
\widehat{\nabla}_{\v\phi_j}\mathcal{L} &= \frac{\partial f(\v\epsilon_j)^\top}{\partial \v\phi_j}\frac{\partial\mathcal{L}}{\partial\v\theta_j} + \underbrace{\cancel{\frac{\partial g(\v\tau_j, \v\epsilon_j)^\top}{\partial\v\phi_j} \frac{\partial\mathcal{L}}{\partial\v z_j}}}_{\text{dependency removed}}, \nonumber\\ 
&= \textcolor{blue}{\frac{\partial f(\v\epsilon_j)^\top}{\partial \v\phi_j}}\bigg [\textcolor{red}{\frac{\partial\mathcal{L}_{0,j}}{\partial\v\theta_j}} + \textcolor{blue}{\frac{\partial\mathcal{L}_{1,j}}{\partial\v\theta_j}} \bigg ]\label{eqn:power_mean_phi_gradient}. 
\end{align}
With the gradients for the variational parameters $\v\phi_j$ and $\v\psi_j$ derived, we now present Algorithm \ref{alg:power_likelihoood} for fitting the power-likelihood model using either the amortized structured or the mean-field variational approximation. The algorithm follows an iterative scheme that requires two communication rounds per iteration. First, the server receives the auxiliary variables $\v z_j$ from each client and sends all clients the complete set of auxiliary variables $\v z$. The clients then compute the gradients of the likelihood with respect to the auxiliary variables of the other clients and send these gradients back to the server. The server aggregates the gradients and sends the respective sums to each client. Finally, the clients update their variational parameters and auxiliary variables locally and send the updated auxiliary variables back to the server. This process is repeated for a specified number of iterations or until convergence.

\begin{algorithm}[ht]
\SetAlgoLined
\KwIn{\textit{Server}: number of iterations $N$. \textit{Clients}: initial local variational parameters $\v\phi_j, \v\psi_j$.}
\KwOut{\textit{Clients}: variational parameters $\v\phi_j, \v\psi_j$.}
\textit{Initialization step:} Each client generates their value for $\v z_j$ and sends it to the server\\
\For{$i=1, \ldots, N$}{
\Server{
Receive $\v z_j$ from client $j=1, \ldots, J$\\
Send $\v z$ to all clients \\

}

\For{each client $j$ in parallel}{
Receive $\v z$ from the server\\
Generate $\v\theta_j$ via \eqref{eqn:f_reparam}\\
Calculate $\big \{\frac{\partial \mathcal{L}_{0, j}}{\partial \v z_k}\big \}_{k\neq j}^J$\\
Send $\big \{\frac{\partial \mathcal{L}_{0, j}}{\partial \v z_k}\big \}_{k\neq j}^J$ to the server}

\Server{
Receive $\big \{\frac{\partial \mathcal{L}_{0, j}}{\partial \v z_k}\big \}_{k\neq j}^J$ for $k=1, \ldots, J$ from the clients\\
Calculate $\sum_{k\neq j}^J \frac{\partial\mathcal{L}_{0, k}}{\partial \v z_j}$ for $j = 1, \ldots, J$\\
Send $\sum_{k\neq j}^J \frac{\partial\mathcal{L}_{0, k}}{\partial \v z_j}$ to the $j$-th client for $j=1, \ldots, J$
}

\For{each client $j$ in parallel}{
Receive $\sum_{k\neq j}^J \frac{\partial\mathcal{L}_{0, k}}{\partial \v z_j}$ from the server\\
Calculate $\widehat{\nabla}_{\v\phi_j}\mathcal{L}$ via \eqref{eqn:power_phi_gradient} for structured or \eqref{eqn:power_mean_phi_gradient} for mean-field\\
Calculate $\widehat{\nabla}_{\v\psi_j}\mathcal{L}$ via \eqref{eqn:power_psi_gradient}\\
$\v\phi_j \gets {\sf optimizer.step}\big (\widehat{\nabla}_{\v\phi_j}\mathcal{L}\big )$\\
$\v\psi_j \gets {\sf optimizer.step}\big (\widehat{\nabla}_{\v\psi_j}\mathcal{L}\big )$\\
Generate $\v z_j$ via \eqref{eqn:g_reparam} for structured or \eqref{eqn:mfvi_reparam} for mean-field\\
Send $\v z_j$ to server

}

}

\caption{SFVI for power-likelihood model}\label{alg:power_likelihoood}
\end{algorithm}

\section{Numerical Examples}\label{sec:numerics}
We provide numerical results for three classes of models. In Section \ref{sec:logistic_regression}, the logistic regression model example showcases the models and variational approximations that we describe and the effect that varying the value of the noise hyperparameter $\rho$ and the number of clients $J$ has on inference. In Section \ref{sec:multilevel_example}, the multilevel Poisson regression model demonstrates the feasibility of fitting complex models while potentially enhancing client privacy through a multi-level structure. In Section \ref{sec:split_learning_example}, we demonstrate a novel hierarchical Bayes split NN model for the task of \emph{split learning}, and compare its' benefits to the originally proposed split NN model.

\subsection{Logistic Regression}\label{sec:logistic_regression}
Here, we fit a logistic regression model and the corresponding augmented-variable and power-likelihood model to a synthetic data set. The descriptions for all three model variations are defined in Table~\ref{tab:logistic_regression_models}.

\begin{table}
    \centering
    \begin{tabular}{lll} \toprule
        Logistic regression model & Augmented-variable model & Power-likelihood model \\ \hline
        $y_i \sim \text{Bernoulli}(p_i)$ & $y_i \sim \text{Bernoulli}(p_i)$ & $y_i \sim \prod_{j=1}^J\text{Bernoulli}(p_{ij})^{1/J}$\\
        $\text{logit}(p_i) = b + \v x_i^\top\v\beta$ & $\text{logit}(p_i) = b + \sum_{j=1}^J z_{ij}$ & $\text{logit}(p_{ij}) = b + \v x_{ij}^\top\v\beta_j + \sum_{m\neq j}^J z_{im}$\\
        & $z_{ij}\sim\mathcal{N}(\v x_{ij}^\top\v \beta_j, \rho)$ & $z_{ij}\sim\mathcal{N}(\v x_{ij}^\top\v \beta_j, \rho)$\\
        $\beta_k \sim \mathcal{N}(0, 1)$ & $\beta_{jk} \sim \mathcal{N}(0, 1)$ & $\beta_{jk} \sim \mathcal{N}(0, 1)$\\ 
        $b \sim \mathcal{N}(0, 1)$ & $b \sim \mathcal{N}(0, 1)$ & $b \sim \mathcal{N}(0, 1)$\\ \bottomrule
    \end{tabular}
    \caption{Comparison of the logistic regression model, augmented-variable model, and power-likelihood model. In all models, $i = 1, \ldots, N$ denotes the observation index. In the logistic regression model, $k = 1, \ldots, p$ denotes the covariate index. In the auxiliary-variable and power-likelihood models, $j = 1, \ldots, J$ denotes the client index, $k = 1, \ldots, p_j$ denotes the covariate index for client $j$, and $m \neq j$ denotes the index for clients other than $j$. The auxiliary-variable $z_{ij}$ is introduced for each observation $i$ and client $j$ in both the auxiliary-variable and power-likelihood models.}
    \label{tab:logistic_regression_models}
\end{table}

We generate the synthetic data from the underlying logistic regression model, with random standard Gaussian noise added to the linear predictor for each observation. The number of observations is $n=500$, and the number of covariates is 
$p=20$. We generate all elements of the data matrix $\v x$ from the standard Gaussian distribution. In this example, we compare the performance of the three models, fit using Markov chain Monte Carlo (MCMC), in particular with the No-U-Turn sampler (NUTS) fit in NumPyro \parencite{phan2019composable}, and two different variational approximations, a mean-field approximation and the amortized approximation. Note that the posterior approximation for the parameters $\v\beta$ and $b$ are the same throughout the two approximations, and the only difference comes in the variational approximation for the parameters $\v z$. Table \ref{tab:logistic_models_algos} shows the combination of models and algorithms that we consider fitting in this example.
\begin{table}
\centering
\begin{tabular}{c ccc}
\toprule
& True model & Augmented-variable & Power-likelihood \\
\midrule
MCMC & \tick & \tick & \tick \\
Mean-field VI & \tick & \tick & \tick \\
Amortized VI & \cross & \tick & \tick \\
\bottomrule
\end{tabular}
\caption{The eight combinations of model and inference algorithms we run. The columns denote the three models: the true model (\ie the true data-generating model), the augmented-variable model, and the power-likelihood model. The rows denote three inference algorithms: MCMC (specifically the No-U-Turn sampler) and two ``federated'' variants, represented as two differing variational approximations, namely mean-field VI }\label{tab:logistic_models_algos}
\end{table}
\begin{figure}[!ht]
    \centering
    \begin{subfigure}[b]{0.96\textwidth}
        \centering
        \includegraphics[width=\textwidth]{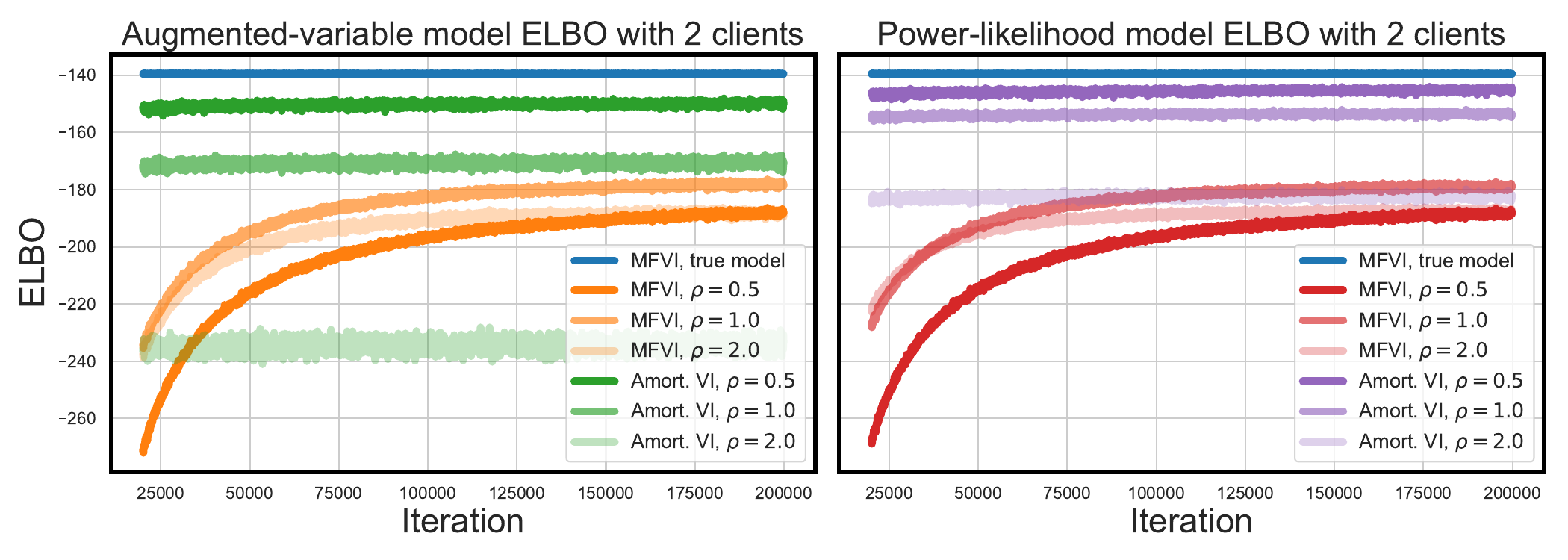}
        \caption{The plot on the left-hand side shows the ELBO values of the variational approximations fit with the augmented-variable model. The plot on the right-hand side shows the performance of the variational approximations with the power-likelihood model. Both plots compare to the ELBO of the true logistic regression model (blue).}
        \label{fig:elbo_two_clients}
    \end{subfigure}
    \\
    \begin{subfigure}[b]{0.96\textwidth}
        \centering
        \includegraphics[width=\textwidth]{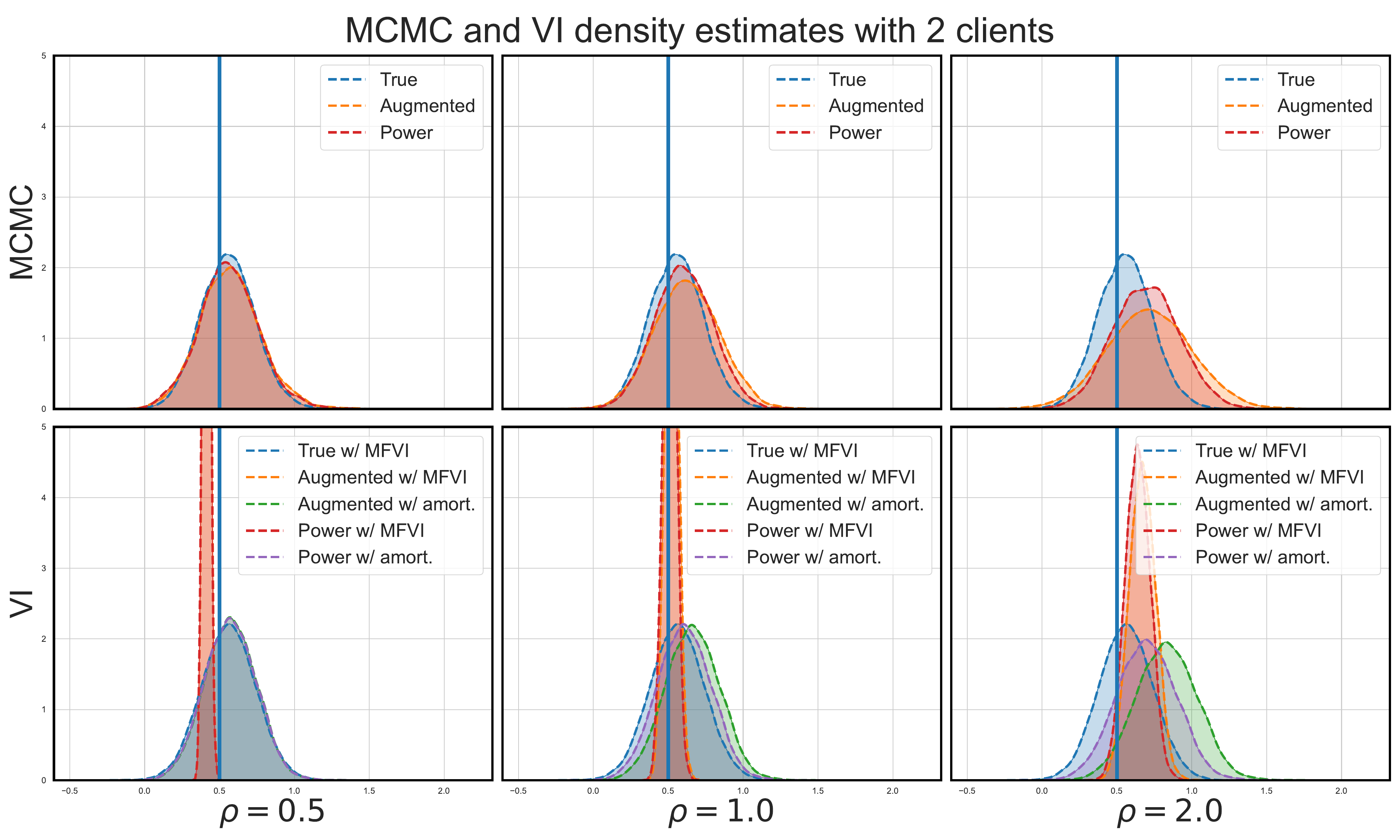}
        \caption{Marginal density plots comparing the performance of the three models for the same parameter. Top row shows MCMC results. Bottom row shows variational approximation results. True parameter value is denoted by the vertical blue line.}
        \label{fig:density_two_clients}
    \end{subfigure}
    \caption{ELBO and marginal density plots for the two-client example. Figure \ref{fig:elbo_two_clients} shows the ELBO values for a combination of models and variational approximations. Figure \ref{fig:density_two_clients} shows marginal densities for the various models using both MCMC and variational approximations.}
    \label{fig:two-client}
\end{figure}

\begin{figure}[!ht]
    \centering
    \begin{subfigure}[b]{0.96\textwidth}
        \centering
        \includegraphics[width=\textwidth]{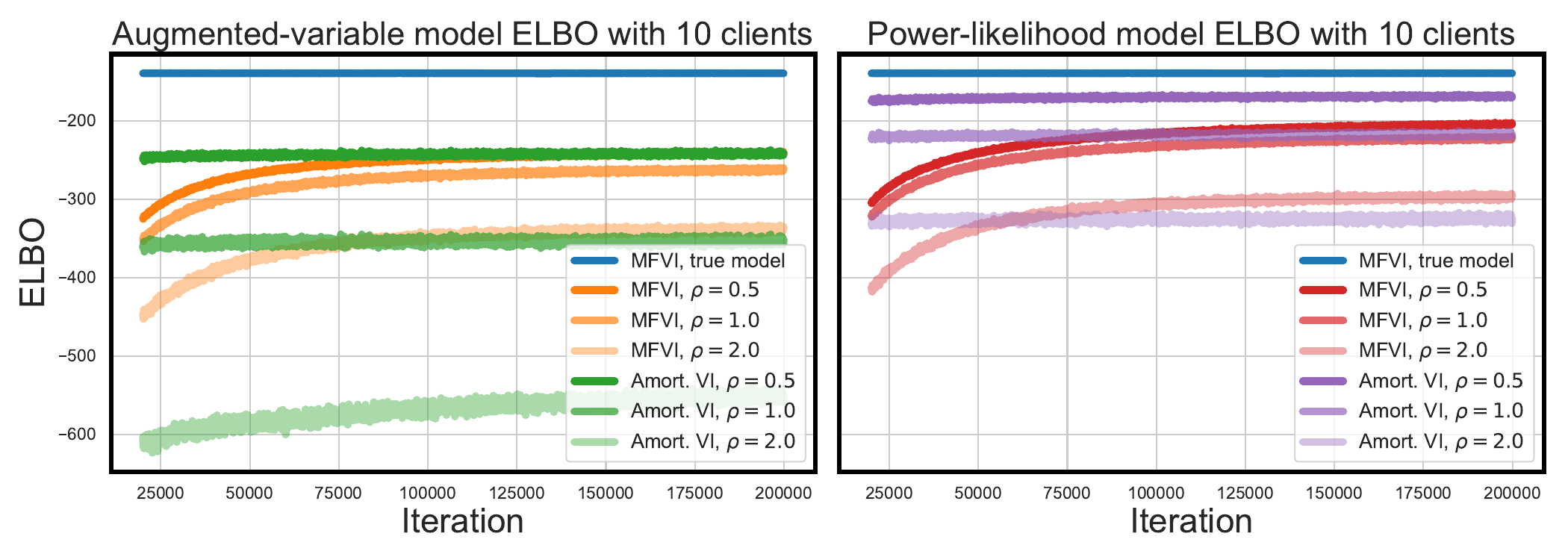}
        \caption{The plot on the left-hand side shows the ELBO values of the variational approximations fit with the augmented-variable model. The plot on the right-hand side shows the performance of the variational approximations with the power-likelihood model. Both plots compare to the ELBO of the true logistic regression model (blue).}
        \label{fig:elbo_ten_clients}
    \end{subfigure}
    \\
    \begin{subfigure}[b]{0.96\textwidth}
        \centering
        \includegraphics[width=\textwidth]{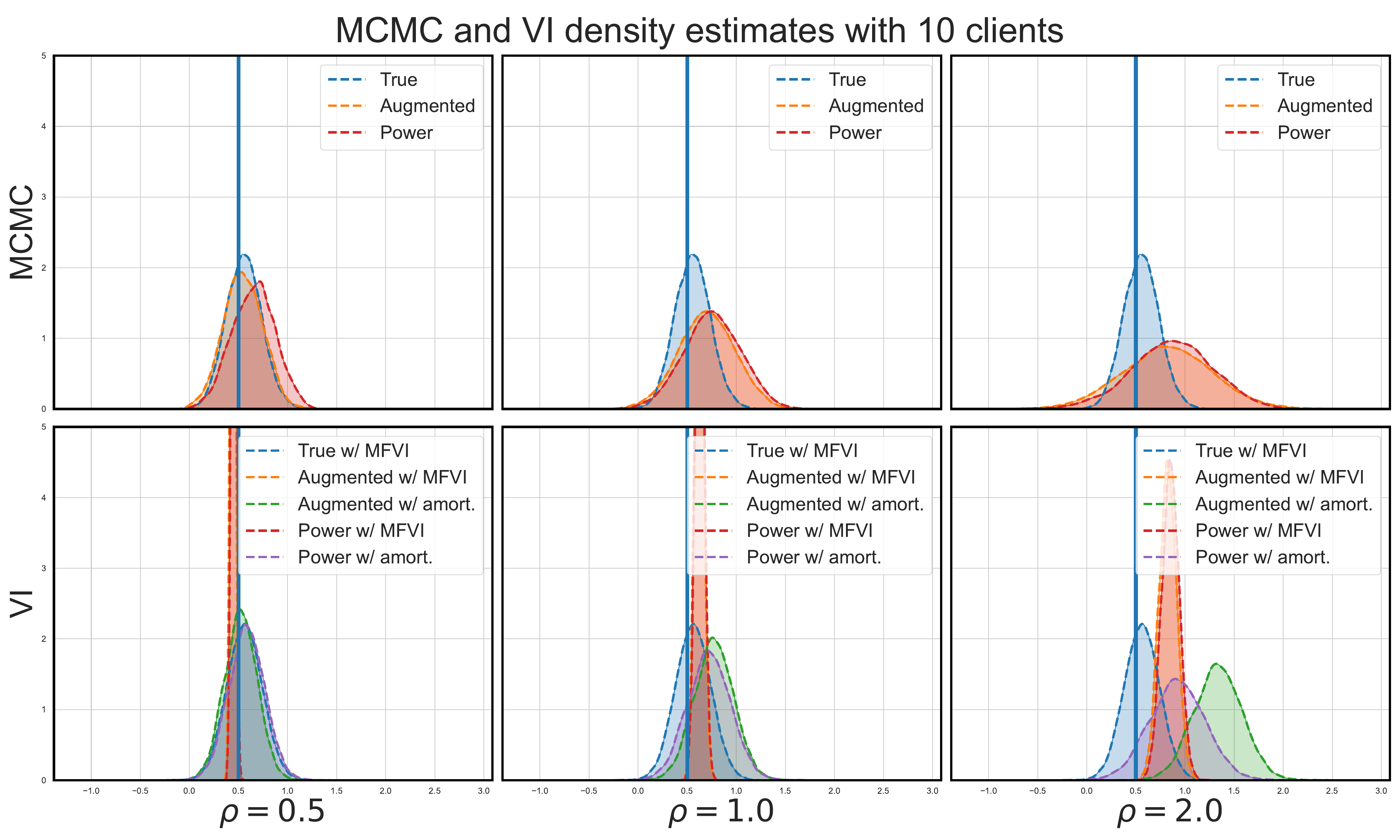}
        \caption{Marginal density plots comparing the performance of the three models for the same parameter. The top row shows MCMC results. The bottom row shows variational approximation results. True parameter value is denoted by the vertical blue line.}
        \label{fig:density_ten_clients}
    \end{subfigure}
    \caption{ELBO and marginal density plots for the ten-client example. Figure \ref{fig:elbo_ten_clients} shows the ELBO values for a combination of models and variational approximations. Figure \ref{fig:density_ten_clients} shows marginal densities for the various models using both MCMC and variational approximations.}
    \label{fig:ten-client}
\end{figure}

We manipulate two key parameters: the noise hyperparameter $\rho$, tested at values ${0.5, 1.0, 2.0}$, and the number of clients, considering scenarios with two clients ($J=2$) holding ten covariates each, and ten clients ($J=10$) with two covariates each. These variations help us explore the models' performance across different federated environments. Figure \ref{fig:two-client} shows the results from the two-client example and Figure \ref{fig:ten-client} shows the results from the ten-client example. 

The ELBO values for the two- and ten-client examples are given by Figures \ref{fig:elbo_two_clients} and \ref{fig:elbo_ten_clients} respectively. The ELBO values show that in both the two- and ten-client examples, the power-likelihood model outperforms the augmented-variable model. There is less of a difference between the two models in the ten-client example compared to the two-client example; this is expected as the weightings of each of the client-specific likelihood contributions reduce with the increase in the number of clients, tending closer towards the augmented-variable model.

Figures \ref{fig:density_two_clients} and \ref{fig:density_ten_clients} show marginal density plots for the same single parameter from MCMC and variational approximations, where the thick blue line represents the true parameter value. We plot the marginal densities of the same, single parameter, as there is no expected difference in quality of the marginal densities across different parameters in the model. For lower values of $\rho$, \ie $\rho=0.5$ or $\rho=1.0$, the amortized approximation tends to outperform MFVI. For high values of $\rho$, \ie $\rho=2.0$, we see that MFVI outperforms the amortized approximation. This is expected as the amortized approximation is estimating a shared function for the marginal posterior approximations for $\v z$, giving differing inputs per observation. As $\rho$ increases, the parameters $\v z$ have greater variance, and it is now difficult to estimate a shared function suitable for all parameters.

\subsection{Poisson Multilevel Regression}\label{sec:multilevel_example}

In this example, we demonstrate the application of a multilevel Poisson regression model to analyze a synthetic dataset designed to emulate the complexities of real-world health data associated with Australian geography at the Statistical Area 2 (SA2) level, which includes approximately $N=2000$ areas. Our analysis focuses on cancer diagnosis counts per area as the response variable $\mathbf{y} \in \mathbb{R}^n$, inspired by the data representation in the \emph{Australian Cancer Atlas} \parencite{duncan2019development}.
We consider a scenario where two clients each hold a unique set of SA2-level health-related covariates, represented as $\v x_1 \in \mathbb{R}^{N \times 2}$ and $\v x_2 \in \mathbb{R}^{N \times 2}$. This setup reflects the real-world challenges of efficiently combining disparate health datasets while addressing logistical and privacy concerns. Additionally, we incorporate two publicly available variables: the population count per SA2 area ($\text{pop}_i$) and a remoteness level indicator ($r_i$) for each area $i$. The log of the population count serves as an offset term to model the rate of cancer diagnoses per area. At the same time, the remoteness indicator allows us to explore how associations vary across different levels of remoteness.

The motivation for this example is twofold. First, we aim to demonstrate the feasibility of fitting multilevel models in the vertical federated learning (VFL) setting using the auxiliary variable approaches described in this paper. Second, fitting multilevel models such as these might enable individual clients to randomly choose the multilevel structure they want to model their data with to hide information about their underlying data while still obtaining meaningful estimates of the covariate effects of interest.

We examine two models in this example: the true underlying Poisson multilevel model from which we simulate the synthetic dataset and the equivalent augmented-variable model. The specification for the Poisson multilevel regression model is
\begin{align*}
y_i &\sim \text{Poisson}(\lambda_i), \quad i=1, \ldots, N,\\
\log (\lambda_i) &= b + \log (\text{pop}_i) + \sum_{j=1}^J \v x_{ij}^\top\v \beta_{ij}^{r_i}, \quad i =1, \ldots, N, \quad j =1, \ldots, J,\\
\v\beta_{ij}^{r_i} &\sim \mathcal{N}(\v\mu_j^{r_i}, \v\sigma_j^{r_i}), \quad j =1, \ldots, J, \\
\v\mu_{j}^{r_i} &\sim \mathcal{N}(0, 1), \quad j=1, \ldots, J, \\
\v\sigma_j^{r_i} &\sim \mathcal{HN}(1), \quad j=1, \ldots, J, \\
b &\sim \mathcal{N}(0, 1).
\end{align*}
The specification for the equivalent augmented-variable model is
\begin{align*}
y_i &\sim \text{Poisson}(\lambda_i), \quad i=1, \ldots, N,\\
\log (\lambda_i) &= b + \log (\text{pop}_i) + \sum_{j=1}^J z_{ij}, \quad i =1, \ldots, N, \quad j =1, \ldots, J,\\
z_{ij} &\sim \mathcal{N}(x_{ij}^\top\v \beta_{ij}^{r_i}, \rho), \quad i=1, \ldots, N, \quad j =1, \ldots, J, \\
\v \beta_{ij}^{r_i} &\sim \mathcal{N}(\v \mu_j^{r_i}, \v \sigma_j^{r_i}), \quad j =1, \ldots, J, \\
\v \mu_{j}^{r_i} &\sim \mathcal{N}(0, 1), \quad j=1, \ldots, J, \\
\v \sigma_j^{r_i} &\sim \mathcal{HN}(1), \quad j=1, \ldots, J, \\
b &\sim \mathcal{N}(0, 1).
\end{align*}

In both model specifications, $i=1, \ldots, N$ denotes the index for the SA2 areas, $j=1, \ldots, J$ denotes the index for the clients, and $r_i$ denotes the remoteness level for area $i$. The term $\v\beta_{ij}^{r_i}$ represents the varying slopes for the covariates held by client $j$ for area $i$, which depend on the remoteness level $r_i$. In the augmented-variable model, $z_{ij}$ denotes the auxiliary variable introduced for area $i$ and client $j$.

This example showcases the augmented-variable model's potential benefits in the context of VFL. By introducing auxiliary variables, the augmented-variable model allows for efficient and privacy-preserving estimation of the multilevel model parameters without requiring direct sharing of the client-specific covariates.

We generate the synthetic data using the true data-generating process of the multilevel Poisson regression model. Each element of the data matrices $\v x_1$ and $\v x_2$ is drawn from a standard Gaussian distribution. The population term is generated from a discrete uniform distribution between the values of $250$ and $350$, reflecting a realistic range of population counts at the SA2 level. The remoteness indicator randomly and uniformly assigns each observation to one of five levels, capturing the variation in remoteness across Australian SA2 regions. After constructing the linear predictor for each observation, we generate the response from a Poisson distribution.

\begin{figure}[!ht]
    \centering
    \includegraphics[width=0.96\textwidth]{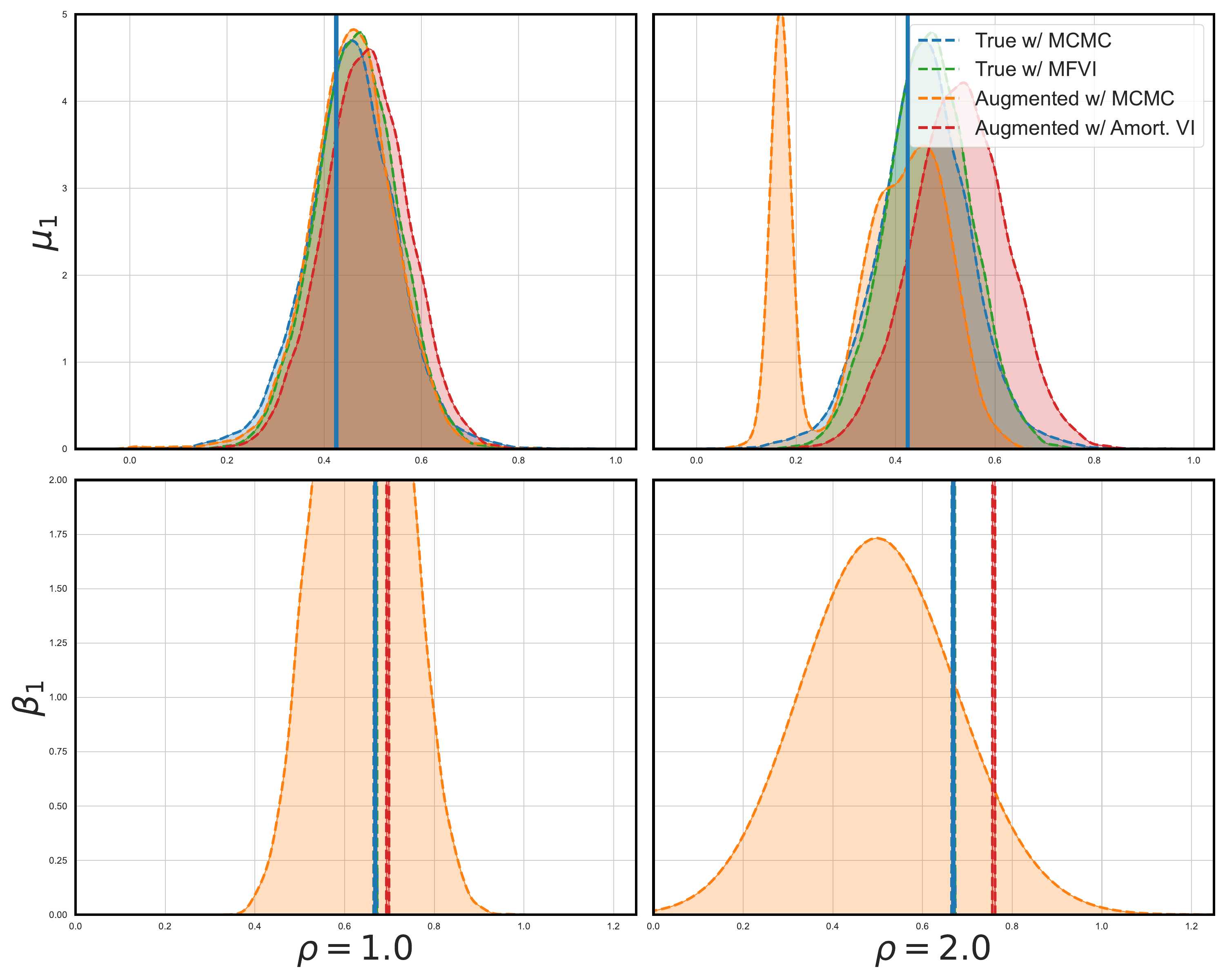}
    \caption{The top row shows posterior density estimates for the global mean effect $\mu_1$ of the first covariate belonging to the first client. The bottom row shows posterior density estimates for one of the $\v\beta_1$ levels for the first covariate on the first client. The left column shows augmented-variable models fit using $\rho=1$, and the right shows augmented-variable models fit using $\rho=2$. The blue line shows the true parameter value.}
    \label{fig:multilevel}
\end{figure}

Figure \ref{fig:multilevel} shows posterior density estimates for the true Poisson multilevel model fit using both MCMC and MFVI, and the augmented-variable model with two different values of $\rho$, fit using MCMC and the amortized structured family proposed in Section \ref{sec:amortized_variational_family}. Posterior density estimates for a single $\mu_1^1$ and $\beta_1^1$ parameter are provided, but we observe similar relationships and patterns for different parameters at the same level. For both $\rho=1$ and $\rho=2$, the augmented-variable model returns reasonable estimates for $\mu_1$ using both MCMC and the amortized structured family, comparable to the true multilevel model fit using either MCMC or MFVI. However, we note that during MCMC sampling of the augmented-variable model, we observe a multimodal density estimate, which signifies potential issues. For the estimate of $\beta_1$, the MCMC of the augmented-variable model yields a wide and uncertain posterior, whereas the amortized VI captures the shape more accurately. It is worth noting that the MCMC and MFVI estimates for $\beta_1$ in the true model are aligned almost precisely with the true value (indicated by the blue line).

Using this model and algorithm approach, the ability to fit arbitrary multilevel structures enables clients to randomly choose the multilevel structure they want to use for modeling the data. This increases privacy, as the values of the auxiliary variables that the clients must communicate depend on a multilevel structure unknown to the other clients. Further exploration of this aspect could help quantify the privacy benefits afforded by this approach.

\subsection{Hierarchical Bayes Split Neural Net}\label{sec:split_learning_example}
Split NN \parencite{poirot2019split, ceballos2020splitnn, thapa2022splitfed} are a specific model developed for vertical FL applications whereby clients may not have access to the response variable; sometimes this task is referred to as \emph{split learning}. In this section, we consider a classification task with two classes. The split NN variant of a logistic regression model is
\begin{align*}
y_i &\sim \text{Bernoulli}(\lambda_i), \quad i = 1, \ldots, N,\\
\text{logit}(\lambda_i) &= \sum_{j=1}^J z_{ij}, \quad i=1, \ldots, N\,\\
z_{ij} &= f_{\v\phi_j}(\v x_{ij}), \quad i = 1, \ldots, N, \quad j=1, \ldots, J, 
\end{align*}
where each $f_{\v\phi_j}$ is a client-specific NN parameterized by $\v\phi_j$, that outputs $\v z_j\in\mathbb{R}^N$. In order to fit the above model via gradient descent using automatic differentiation, updating the parameters $\v\phi_j$ requires each client to calculate $\frac{\partial p(y_i | \lambda_i)}{\partial z_{ij}}\frac{\partial z_{ij}}{\partial\v\phi_j}$. If each client communicates $z_{ij}$ to the server, the server can calculate and return $\frac{\partial p(y_i | \lambda_i)}{\partial z_{ij}}$ to each client. Each client can independently calculate $\frac{\partial h_{ij}}{\partial\v\phi_j}$
and thus have the necessary gradient information to calculate the required update.

\begin{figure}[!ht]
    \centering
    \begin{subfigure}[b]{0.48\textwidth} 
        \centering
        \includegraphics[width=\textwidth]{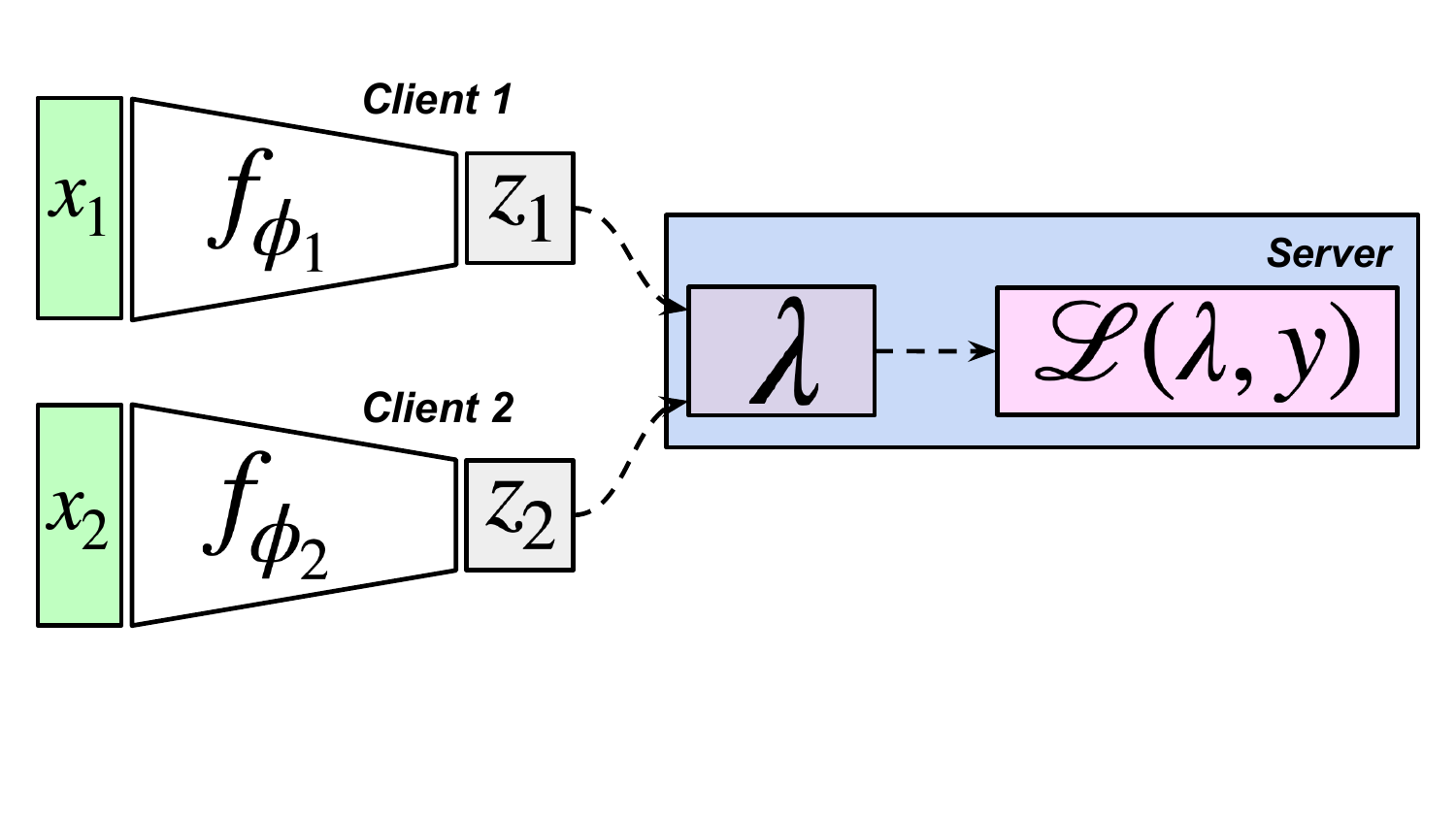}
        \caption{Split NN}
        \label{fig:split_NN}
    \end{subfigure}
    ~ 
    \begin{subfigure}[b]{0.48\textwidth} 
        \centering
        \includegraphics[width=\textwidth]{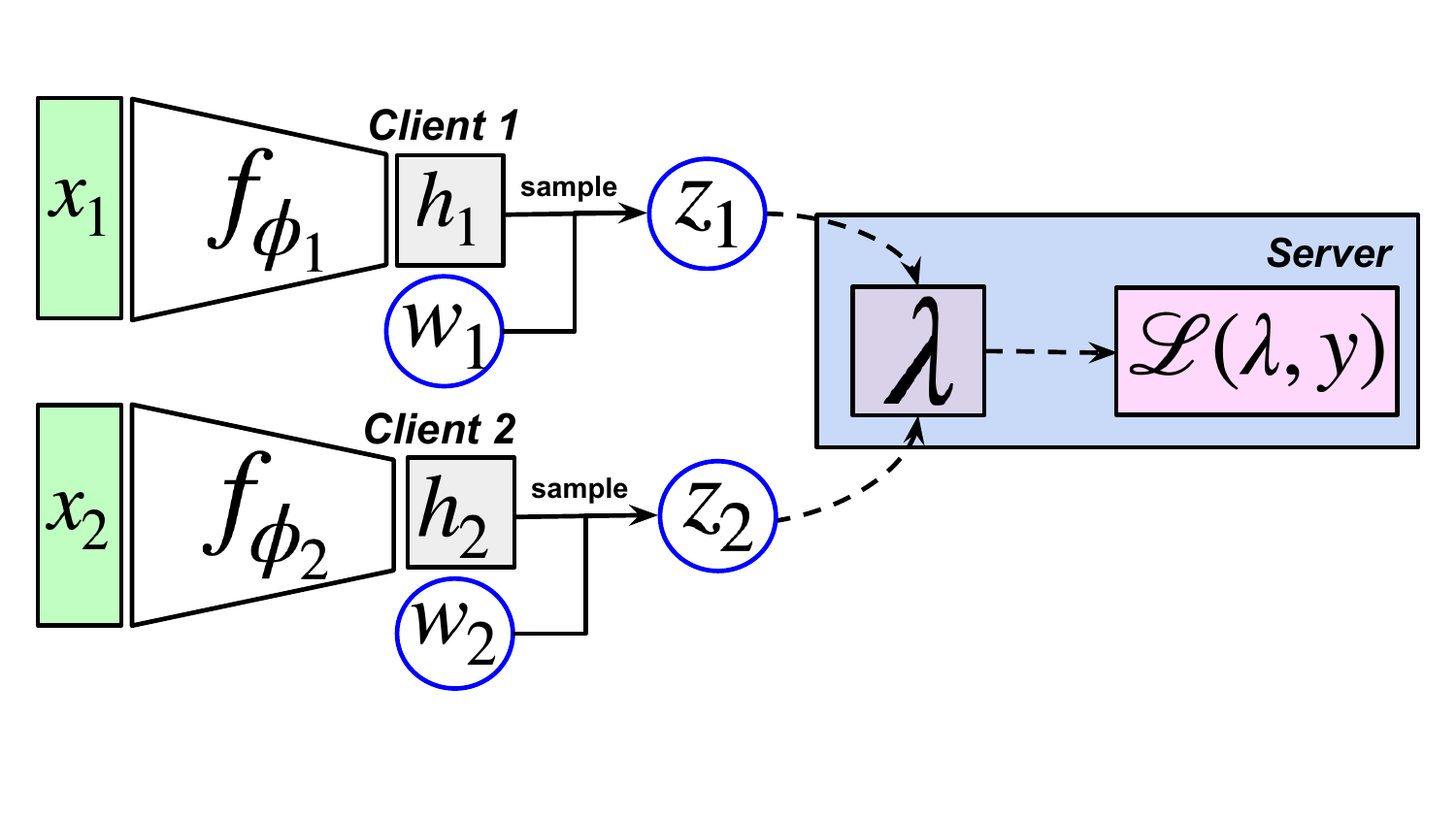}
        \caption{Hierarchical Bayes Split NN}
        \label{fig:hier_Bayes_split_NN}
    \end{subfigure}
    \caption{Figure \ref{fig:split_NN} shows the \emph{split NN} model. Figure \ref{fig:hier_Bayes_split_NN} shows our proposed \emph{hierarchical Bayes split NN}. In the split NN, each client learns a function $f_{\v\phi_j}$, parameterized as a NN with weights $\v\phi_j$, that maps the covariates from each client to a vector $\v z_j \in\mathbb{R}^N$. Each client communicates their $\v z_j$ to the server, where the set of $\v z_j$'s is used to create the linear predictor $\v\lambda$, which is then used to evaluate the objective function, \ie likelihood, on the server. In the hierarchical Bayes variant, the final-layer weights of the function $f_{\v\phi_j}$ 
    are denoted as $\v w_j$. These weights are treated as random variables and are assigned a prior distribution. We dot-product the weights with the hidden vector $\v h_j$ to parameterize the mean of the prior distribution of an additional set of random variables $\v z_j$. Each client sends their respective parameters $\v z_j$ to the server, and the server takes the same steps as before.}
    \label{fig:split_example_diagram}
\end{figure}

Using the augmented-variable model and the amortized variational approximation, we develop a hierarchical Bayes variant of the split NN as follows,
\begin{align*}
y_i &\sim \text{Bernoulli}(\lambda_i), \quad i = 1, \ldots, N,\\
\text{logit}(\lambda_i) &= \sum_{j=1}^J z_{ij}, \quad i=1, \ldots, N,\\
z_{ij} &\sim \mathcal{N}(\v h_{ij}^\top \v w_j, \rho),  \quad i = 1, \ldots, N, \quad j=1, \ldots, J, \\
\v h_{ij} &= f_{\v\phi_j}(\v x_{ij}), \quad i = 1, \ldots, N, \quad j=1, \ldots, J, \\
\v w_j &\sim \mathcal{N}(0, 1), \quad j = 1, \ldots, J.
\end{align*}
In this hierarchical Bayes variant, the final-layer weights $\v w_j$ of the function $f_{\v\phi_j}$ are treated as random variables and assigned a prior distribution. The dot product the hidden vector 
$\v h_{ij}$ with the weights $\v w_j$ parameterizes the mean of the prior distribution for auxiliary variable $\v z_{ij}$. Each client sends their respective parameters $\v z_j$ to the server, and the server performs the same steps as in the standard split NN. Figure \ref{fig:split_example_diagram} illustrates the differences between the formulations of the \emph{hierarchical Bayes split NN} and the original \emph{split NN}.

Bayesian models offer advantages such as natural probabilistic interpretations of uncertainty and more accurately calibrated estimates, especially when trained on limited data. In this FL context, the Bayesian analog of split 
NNs provides two specific benefits related to privacy and efficient computation. First, the random variable $z_{ij}$ has a Gaussian prior distribution with a standard deviation equal to the hyperparameter $\rho$; this allows the client to send a noisy estimate of the function output of $f_{\v\phi_j}$ to the server, enhancing privacy. Second, the hierarchical Bayes variant offers computational advantages. In a split NN, only one gradient step can be taken for the weights $\v\phi_j$ per communication round with the server, as the value $\frac{\partial p(y_i | \lambda_i)}{\partial z_{ij}}$ is required at each step. The hierarchical Bayes variant benefits from $z_{ij}$ being a stochastic function of $x_{ij}$ and the conditional independence of the parameters $\v\phi_j$ and local latent variables from the likelihood function given $\v z_j$. This allows for multiple updates of $\v\phi_j$ and $\v w_j$ for a fixed value of $\v z_j$ without requiring communication with the server.

We evaluate the performance of the hierarchical Bayes split NN using the heart disease prediction dataset from Kaggle \parencite{fedesoriano_2021}. This dataset, created by aggregating five existing datasets, contains $918$ unique patients (\ie observations) with $11$ covariates and a binary response indicating the presence of heart disease. We preprocess the data by normalizing all continuous variables via a Z-score transform and creating one-hot-encodings for all categorical variables. The dataset is split between two clients, with client $1$ allocated the first five covariates and client $2$ allocated the remaining six covariates. 

To assess the model's performance, we employ a 10-fold cross-validation approach, splitting the dataset into $10$ different 90\%/10\% training and test splits. The model is trained and evaluated on each split, ensuring that all observations appear in a test set exactly once. We use a standard $2$-layer feedforward NN architecture with $8$ dimensions for the weights in the first two layers and $2$ dimensions for the weights in the final layer. All models are trained for $50,000$ iterations using the Adam optimizer \parencite{kingma2014adam} with a learning rate of $1e^{-3}$.

\begin{table}[htbp]
\centering
\begin{tabular}{cccc}
\toprule
Model & \begin{tabular}{@{}c@{}}Prediction\\accuracy (\%)\\per batch\end{tabular} & \begin{tabular}{@{}c@{}}Average log-\\likelihood\\(all predictions)\end{tabular} & \begin{tabular}{@{}c@{}}Average log-\\likelihood\\(incorrect predictions)\end{tabular} \\
\midrule
Split NN & \textcolor{red}{$\mathbf{80.73 \% \pm 3.65}$} & \textcolor{red}{$\mathbf{-2.21 \pm 4.41}$} & \textcolor{red}{$\mathbf{-10.95 \pm 1.99}$} \\
\begin{tabular}{@{}c@{}}Hier. Bayes split NN, $\rho=1$\end{tabular} & $\mathbf{86.39\% \pm 4.20}$ & $\mathbf{-0.36 \pm 0.84}$ & $\mathbf{-1.99 \pm 1.42}$ \\
\begin{tabular}{@{}c@{}}Hier. Bayes split NN, $\rho=5$\end{tabular} & $84.98\% \pm 4.92$ & $-0.59 \pm 1.71$ & $-3.53 \pm 3.01$ \\
\begin{tabular}{@{}c@{}}Hier. Bayes split NN, $\rho=10$\end{tabular} & $84.10\% \pm 5.23$ & $-0.67 \pm 1.99$ & $-3.89 \pm 3.53$ \\
\bottomrule
\end{tabular}
\caption{Comparison of the test predictive accuracy and average test log-likelihood for the split NN versus the hierarchical Bayes split NN.}
\label{tab:split_nn_results}
\end{table}

Table \ref{tab:split_nn_results} presents the test predictive accuracy and average test log-likelihood for the split NN and the hierarchical Bayes split NN with different values of the hyperparameter $\rho$ ($\rho = {1, 5, 10}$). The results demonstrate that all variants of the hierarchical Bayes split NN outperform the standard split NN, with better performance observed for smaller values of $\rho$, which is expected, because using higher values of $\rho$ is analogous to reducing the amount of signal that the set of auxiliary variables contains regarding the mean, which is the value of interest. Notably, higher variance is observed for larger values of $\rho$, which is intuitive as it introduces additional variance to the model.

It is worth acknowledging that the split NN exhibits signs of overfitting. Various strategies, such as sub-sampling, dropout \parencite{baldi2013understanding, gal2016dropout}, or additional regularization terms in the objective function, can mitigate overfitting. While the Bayesian variant is closely associated with regularization, the above-mentioned strategies are equally applicable to both standard and Bayesian NNs. In this study, we compare the Bayesian and non-Bayesian variants using a standard architecture and objective function to ensure a fair comparison.

The performance of the hierarchical Bayes split NN supports the growing evidence that Bayesian NNs with stochastic last layers can provide improved results \parencite{kristiadi2020being, sharma2023bayesian, osband2024epistemic, harrison2024variational}. This numerical section demonstrates the effectiveness of the proposed hierarchical Bayes split NN in the context of VFL, highlighting its advantages in terms of privacy and computational efficiency. Future research could explore the application of this approach to other datasets and investigate the impact of different model architectures and hyperparameter settings on performance.

\section{Discussion}\label{sec:discussion}
This paper represents, to our knowledge, the first contribution to the literature on Bayesian methods in the VFL setting. We formulated auxiliary variable methods for fitting Bayesian models in VFL and designed a new model type tailored for specific VFL settings, demonstrating improved performance over existing augmented-variable models. The auxiliary variable methods used in this paper increase the dimensionality of the model, with the number of additional parameters equaling the product of the number of observations and clients. To address this, we propose a factorized amortized approximation that prevents the dimensionality of the inference task from scaling with the number of observations while improving performance over standard Gaussian variational approximations, exhibiting faster convergence and enabling subsampling.

We showcase the proposed methods through three illustrative examples. First, we fit a logistic regression model and compare the estimates from the actual, augmented-variable, and novel power-likelihood models using both MCMC and variational approximations, assessing performance with varying values of the privacy-aiding hyperparameter $\rho$ and a varying number of clients. Second, we fit a complex multilevel regression model, demonstrating the ease of fitting multilevel models and the ability of clients to create an additional multilevel structure unknown to other clients or the server, potentially enhancing data privacy while maintaining accurate parameter estimates. Lastly, we introduce a novel hierarchical Bayes split NN and compare it to a split NN, showcasing the scalability of the proposed Bayesian auxiliary variable methods with the amortized approximation, handling large numbers of observations and incorporating sub-sampling. The hierarchical Bayes variant exhibits similar test accuracy but a higher average test log-likelihood than the original, indicating better-calibrated estimates. Additional benefits of the hierarchical Bayes variant include the addition of the $\rho$ hyperparameter, aiding privacy and enabling more computationally and communication-efficient algorithmic schemes by exploiting the conditional independence created.

This work opens up numerous promising avenues for future investigation. While the current algorithms are not embarrassingly parallel for updating the parameters $\v z$, specific schemes like the SOUL algorithm \parencite{de2021efficient}, for which we provide derivations in Appendix \ref{appendix:SOUL}, or other EM-style algorithms \parencite{kuntz2023particle, sharrock2023coinem, gruffaz2024stochastic} can fit the same models with fewer iterations for estimating $\v z$. However, the SOUL algorithm provides only maximum a posteriori (MAP) estimates for $\v z$, resulting in narrower posterior distributions for $\v\theta$ conditional on $\v z$ rather than marginalizing over $\v z$. Moreover, using algorithms like SOUL would make the inference problem's scaling dependent on the dimensionality of $\v z$, potentially causing issues as SOUL was initially used in low-dimensional $\v z$ scenarios.

Another promising research direction involves developing mechanisms to estimate weight parameters for each client's likelihood contributions, aiding model selection and ranking the importance of variables and clients. Asynchronous client updates, an open question in most distributed Bayesian inference settings \parencite{winter2024emerging}, represent another avenue to elevate the framework's practical utility. In contrast to the conventional synchronous approach, where client updates are coordinated, leading to bottlenecks as the client count increases, asynchronous updates allow clients to submit updates independently without strict synchronization demands. Asynchronous updates offer reduced latency, increased resilience, and enhanced scalability, making the proposed methods more suitable for real-world applications and industry settings.

\printbibliography

\end{refsection}

\appendix
\section{Derivations for the SOUL algorithm}\label{appendix:SOUL}
\begin{refsection}
Here we describe how the models formulated in Sections \ref{subsec:augmented_model} and \ref{subsec:power_likelihood} can be inferred in the VFL setting using the \emph{stochastic optimization via unadjusted Langevin} (SOUL)\parencite{de2021efficient} algorithm or its federated variant \parencite{kotelevskii2022fedpop} by exploiting the conditional independence structure that we induce among the clients, given $\v z$. 

\subsection{Stochastic Optimization via Unadjusted Langevin}
An alternative to using MCMC or variational approximation algorithms are maximum marginal likelihood estimation (MMLE) algorithms. One might view the use of MMLE as an \emph{empirical Bayes} \parencite{casella1985introduction, carlin2000empirical} procedure, where the auxiliary variables $\v z$ are hyperparameters, and we find the \emph{maximum a posteriori} (MAP) estimates $\hat{\v z}$. The SOUL algorithm results in MAP estimates for $\hat{\v z}$, and draws from the \emph{conditional} target density $p(\v \theta | \hat{\v z}, \v y; \rho)$.

If the dimension is large, the optimization of the hyperparameters can be achieved using a \emph{Robbins-Monro} scheme \parencite{robbins1951stochastic, delyon1999convergence}. 

In the VFL setting with $J$ clients, parameters $\v\theta=(\v\theta_1^\top, \ldots, \v\theta_J^\top)^\top$, where $\v\theta_j\in\mathbb{R}^{p_j}$, auxiliary variables $\v z = (\v z_1^\top, \ldots, \v z_J^\top)^\top$, where $\v z_j \in\mathbb{R}^{n}$, the gradient with respect to $\v z$ of the log marginal likelihood conditional on $\v z$ and the hyperparameter $\rho$ is written as
\begin{align*}
\nabla_{\v z_j} \log p(\v y | \v z; \rho) &= \int \frac{\nabla_{\v z_j}p(\v \theta, \v y | \v z; \rho)}{p(\v\theta, \v y | \v z; \rho)}p(\v \theta | \v y, \v z; \rho)d\v \theta, \\
&= \int \nabla_{\v z_j} \log p(\v\theta, \v y | \v z; \rho)p(\v \theta| \v y, \v z; \rho)d\v\theta, 
\end{align*}
allowing us to calculate an estimator $\widehat{\nabla}_{\v z_j}\log p(\v y | \v z , \rho )$ for the gradient as, 
\begin{align*}
\widehat{\nabla}_{\v z_j}\log p(\v y | \v z , \rho) = \frac{1}{m}\sum_{m=1}^M \nabla_{\v z_j}\log p(\v\theta_k^{(m)}, \v y | \v z; \rho), \quad \v\theta^{(m)}\sim p(\v\theta_j | \v y, \v z; \rho), \quad m=1, \ldots, M.
\end{align*}
\subsubsection{Augmented-variable model details}
Due to the formulation of the augmented target density in Section \ref{subsec:augmented_model}, we can write the log conditional density of $\v\theta_j$, independent of other elements $\v\theta_{-j}$, as
\begin{align}
p(\v\theta_j | \v y, \v z; \rho) \propto \exp\big ( - \log p(\v z_j | \v\theta_j; \rho) - \log p(\v\theta_j)\big ), 
\end{align}
and the gradient 
as 
\begin{align}
\nabla_{\v z_j}\log p(\v\theta, \v y | \v z; \rho) \propto \nabla_{\v z_j}\big [\exp\big (- \log p(\v y|\v z) - p(\v z_j | \v\theta_j; \rho)\big )\big ]. 
\end{align}
For each client $j$, both the above terms are independent of $\v\theta_{-j}$, enabling each client to draw new values of $\v\theta_j$ and update their estimate for $\hat{\v z}_j$. Each client communicate their new estimate of $\hat{\v z}_j$ to the server, that can then calculate and returns the value $\sum_{j=1}^J \hat{\v z}_j$ to each client. The algorithm can iterate in this manner until convergence or until reaching a stopping condition. 

\subsection{Power-likelihood model details}

We demonstrate that the power-likelihood model proposed in Section \ref{subsec:power_likelihood} integration of the proposed \emph{power -likelihood} model can be fit with the SOUL algorithm.
Recall that the log density of the power-likelihood model, conditional on $\v z$ is equal to
\begin{align*}
\log p(\v y, \v\theta | \v z; \rho) = \frac{1}{J}\sum_{j=1}^J \log p(\v y |\v\theta_j, \v z_{-j}) + \sum_{j=1}^J \log p(\v z_{j}|\v\theta_j; \rho)+\sum_{j=1}^J \log p(\v\theta_j).
\end{align*}

The gradient terms 
with respect to $\v\theta_j$ 
required for drawing new values of $\v\theta_j$ via an unadjusted Langevin algorithm are
\begin{align*}
\nabla_{\v\theta_j}\log p(\v y, \v\theta|\v z; \rho) = \nabla_{\v \theta_j}\bigg [\frac{1}{J}\log p(\v y|\v\theta_j, \v z_{-j}; \rho) + \log p(\v z_{j}|\v\theta_j; \rho)+\log p(\v\theta_j) \bigg]. 
\end{align*} 
These gradients are conditionally independent across the $J$ clients given $\v z$, allowing the unadjusted Langevin steps to run in parallel across the $J$ clients. However, added complexity arises when considering the gradient with respect to $\v z_j$, which is equal to
\begin{align}\label{eqn:power_z_full_grad}
\nabla_{\v z_j}\log p(\v y | \v z; \rho) &= \int \nabla_{\v z_j}\log p(\v y, \v\theta|\v z; \rho) p(\v\theta | \v y, \v z; \rho)d\v\theta , \\&= \int\nabla_{\v z_{j}}\bigg [\frac{1}{J}\color{red}\sum_{k\neq j}^J \log p(\v y|\v\theta_k, \v z_{-k}; \rho)\color{black} + \log p(\v z_j|\v\theta_j; \rho) \bigg]p(\v\theta|\v y, \v z; \rho)d\v\theta, \\
&=\mathbb{E}_{p(\v\theta | \v y, \v z; \rho)}\bigg [\nabla_{\v z_j}\bigg [\frac{1}{J}\color{red}\sum_{k\neq j}^J\log p(\v y | \v\theta_k, \v z_{-k}; \rho) \color{black}+ \log p(\v z_j | \v \theta_j; \rho)\bigg ]\bigg ], \\
&= \mathbb{E}_{p(\v\theta_j | \v y, \v z; \rho)}\bigg [\nabla_{\v z_j}\bigg [ \log p(\v z_j | \v \theta_j; \rho)\bigg ]\bigg ]\\ 
&\qquad + \sum_{k\neq j}^J
\mathbb{E}_{p(\v\theta_k | \v y, \v z; \rho)}\bigg [\nabla_{z_j}\bigg [\frac{1}{J}\color{red}\log p(\v y | \v\theta_k, \v z_{-k}; \rho) \color{black}\bigg ]\bigg ].
\end{align}

Here, the challenge is highlighted by the terms that are \color{red} red \color{black} in colour, indicating densities dependent on terms unique to other clients. This dependency implies that updating $\v z_j$ on each client, independently and in parallel, is not feasible. Therefore, after running ULA to obtain $M$ draws, $\{\v \theta_j^{(m)}\}_{m=1}^M$, client $j$ computes the following $J$ terms,
\begin{equation*}
  \text{Client } j \text{ computes }= 
  \begin{cases} 
    \sum_{m=1}^M \nabla_{\v z_{k}}\log p(\v y , \v\theta_j^{(m)}, \v z_{k}), \quad \text{required for update of } \v z_{k} \text{ where } k\neq j, \\
    \sum_{m = 1}^M \nabla_{\v z_{j}}\log p(\v z_{ j}|\v\theta_j^{(m)}; \rho), \quad \text{required for update of } \v z_{j},
    \end{cases}
\end{equation*}
enabling the server to calculate the gradient in \eqref{eqn:power_z_full_grad}, and update the vector of variables $\v z$ on the server, and sending each set of updated variables $\v z$ to each client.
\subsection{Shared parameter details}
Assume that we now want the expression $\nabla_{\sigma}p(\v y | \v z, \sigma)$, where 
\begin{align}
    \nabla_{\sigma}p(\v y | \v z, \sigma) &= \int \nabla_\sigma \log p(\v y, \v\theta|\v z, \sigma; \rho)p(\v\theta|\v y, \v z, \sigma; \rho)d\v\theta, \\
    &= \nabla_{\sigma}\big [\log p(\v y | \v z, \sigma) + \log p(\sigma)\big ].
\end{align}
If all of the values of $\v z$ are communicated on the server, then we can run multiple steps of the gradient descent using the above expression on the server, and then communicate the values of $\v z$ and $\sigma$. For the augmented power likelihood model, the gradient is
\begin{align}
\nabla_{\sigma}p(\v y | \v z, \sigma) &= \int \nabla_{\sigma}\log p(\v y, \v\theta | \v z, \sigma; \rho)p(\v\theta | \v y, \v z, \sigma; \rho)d\v\theta, \\
&= \mathbb{E}_{p(\v\theta | \v y, \v z, \sigma; \rho)}\big [\nabla_\sigma [\log p(\v y | \v \theta, \v z,  \sigma) + \log p(\sigma)]\big ], \\
&= \mathbb{E}_{p(\v\theta | \v y, \v z, \sigma; \rho)}\bigg [\nabla_\sigma \big [\frac{1}{J}\sum_{j=1}^J \log p(\v y | \v \theta_j, \v z_{-j},  \sigma) + \log p(\sigma)\big ]\bigg ],\\
&= \nabla_\sigma \log p(\sigma) + \frac{1}{J}\sum_{j=1}^J \mathbb{E}_{p(\v\theta_j | \v y, \v z, \sigma; \rho)}\big [\nabla_\sigma[\log p(\v y | \v\theta_j, \v z_{-j}, \sigma)]\big ].
\end{align}
If each client $j$, communicates $\mathbb{E}_{p(\v\theta_j | \v y, \v z, \sigma; \rho)}\big [\nabla_\sigma[\log p(\v y | \v\theta_j, \v z_{-j}, \sigma)]$ to the server, then the server can update the value of $\sigma$.
\printbibliography[heading=subbibintoc, title={Appendix References}]
\end{refsection}
\end{document}